\documentclass[
  journal=pasa,
  manuscript=research-paper,
  year=2024,
  volume=37,
]{cup-journal}

\usepackage{amsmath}
\usepackage[nopatch]{microtype}
\usepackage{graphicx}	
\usepackage{amsmath}	
\usepackage{amssymb}	
\usepackage{booktabs,multirow,threeparttable}
\usepackage{color,dsfont,tikz}
\usepackage{orcidlink}
\usepackage{natbib}

\newcommand{\sami}[1]{{SAMI}#1}
\newcommand{\illustristng}[1]{{IllustrisTNG}#1}
\newcommand{\eagle}[1]{{EAGLE}#1}
\newcommand{\simspin}[1]{\textsc{SimSpin}#1}
\newcommand{\reff}[1]{$R_{\text{eff}}$#1}

\newcommand{\sersic}[1]{S\'{e}rsic{#1}}
\def\checkmark{\tikz\fill[scale=0.4](0,.35) -- (.25,0) -- (1,.7) -- (.25,.15) -- cycle;} 

\title[Galaxy Shape Recovery using MDN]{Galaxy 3D Shape Recovery using Mixture Density Network} 

\author{Suk Yee Yong \orcidlink{0000-0002-5204-2902}}
\affiliation{CSIRO Space and Astronomy, PO Box 76, Epping, NSW 1710, Australia}
\alsoaffiliation{ARC Centre of Excellence for All Sky Astrophysics in 3 Dimensions (ASTRO 3D)}
\email[S. Yong]{\href{mailto:sukyee.yong@csiro.au}{sukyee.yong@csiro.au}}

\author{K. E. Harborne \orcidlink{0000-0002-2043-7985}}
\affiliation{International Centre for Radio Astronomy (ICRAR), M468, The University of Western Australia, 35 Stirling Highway, Crawley, WA 6009, Australia}
\alsoaffiliation{ARC Centre of Excellence for All Sky Astrophysics in 3 Dimensions (ASTRO 3D)}

\author{Caroline Foster \orcidlink{0000-0003-0247-1204}}
\affiliation{School of Physics, University of New South Wales, Sydney, NSW 2052, Australia}
\alsoaffiliation{ARC Centre of Excellence for All Sky Astrophysics in 3 Dimensions (ASTRO 3D)}

\author{Robert Bassett}
\affiliation{The Centre for Astrophysics \& Supercomputing, Swinburne University of Technology, P.O. Box 218, Hawthorn, VIC 3122, Australia}
\alsoaffiliation{ARC Centre of Excellence for All Sky Astrophysics in 3 Dimensions (ASTRO 3D)}

\author{Gregory B. Poole}
\affiliation{Astronomy Data and Computing Services (ADACS), Swinburne University of Technology, P.O. Box 218, Hawthorn, VIC 3122, Australia}
\alsoaffiliation{The Centre for Astrophysics \& Supercomputing, Swinburne University of Technology, P.O. Box 218, Hawthorn, VIC 3122, Australia}

\author{Mitchell Cavanagh}
\affiliation{International Centre for Radio Astronomy (ICRAR), M468, The University of Western Australia, 35 Stirling Highway, Crawley, WA 6009, Australia}


\begin{document}

\begin{abstract}
Since the turn of the century, astronomers have been exploiting the rich information afforded by combining stellar kinematic maps and imaging in an attempt to recover the intrinsic, three-dimensional (3D) shape of a galaxy. A common intrinsic shape recovery method relies on an expected monotonic relationship between the intrinsic misalignment of the kinematic and morphological axes and the triaxiality parameter. Recent studies have, however, cast doubt about underlying assumptions relating shape and intrinsic kinematic misalignment. In this work, we aim to recover the 3D shape of \textit{individual} galaxies using their projected stellar kinematic and flux distributions using a supervised machine learning approach with mixture density network (MDN). Using a mock dataset of the \eagle{} hydrodynamical cosmological simulation, we train the MDN model for a carefully selected set of common kinematic and photometric parameters. Compared to previous methods, we demonstrate potential improvements achieved with the MDN model to retrieve the 3D galaxy shape along with the uncertainties, especially for prolate and triaxial systems. We make specific recommendations for recovering galaxy intrinsic shapes relevant for current and future integral field spectroscopic galaxy surveys.
\end{abstract}

\section{Introduction} \label{sec:intro}

The intrinsic, three-dimensional (3D) shape of a galaxy is a fundamental property that has been shown to closely relate to a variety of other physical parameters \citep[e.g.][]{Ryden2006IntrinsicShapeAtlas, Weijmans2014ShapesOfEarlyTypes, vandeSande2018StellarAgesShapes}. 
Indeed, projected visual shape has been a key discriminator of galaxy type since the early classification schemes of \citet{Hubble1926}. 
A key challenge in linking observed galaxy shapes with other physical properties, however, is the reliable translation of two-dimensional (2D) projected information into intrinsic shape properties \citep[e.g.][]{Padilla2008ShapesInSloan, Weijmans2014ShapesOfEarlyTypes, Foster2016SluggsStellarKinematics}. 
Under the simplifying assumption that all galaxies can be reasonably approximated as 3D ellipsoids, a galaxy with any particular 3D shape can be projected into a wide range of 2D elliptical shapes. 
A perfectly axisymmetric disc galaxy, for example, may appear as circular when viewed face-on, or highly elliptical when viewed edge-on (with the axis ratio limited only by the disc's intrinsic thickness).
Due to this degeneracy, the problem of inferring the intrinsic shape of galaxies has long vexed astronomers.

Throughout this work, and as is common in the literature \citep[e.g.][]{Contopoulos1956, Binggeli1980,Franx1991OrderedNatureEllipticals, Lambas1992,Weijmans2014ShapesOfEarlyTypes,Foster2017SAMIIntrinsicShapes}, we assume that a galaxy's 3D shape can be described by an ellipsoid and that its 2D projection can similarly be estimated as an ellipse. 
The projected ellipse shape can be fully described with the lengths of its principal axes as follows:
\begin{equation}
  \frac{x^{2}}{A^{2}} + \frac{y^{2}}{B^{2}} = 1,
\end{equation}
where the semi-major and semi-minor axes, $A$ and $B$, are aligned with the x- and y-directions, respectively. Similarly, an ellipsoid can be fully described as follows:
\begin{equation}
  \frac{x^{2}}{a^{2}}+\frac{y^{2}}{b^{2}}+\frac{z^{2}}{c^{2}} = 1,
\end{equation}
where $a$, $b$, and $c$ are the lengths of the principal semi-axes aligned with the associated Cartesian directions ($x, y, z$) such that $a > b > c$. 
In galaxy shape studies, we are typically interested in the axis ratios rather than the absolute length of any axis and can thus reduce the parameterisation of shape down to one or two values for an ellipse and an ellipsoid, respectively. 
Without loss of generalisation, we may thus quantify 2D shapes through the axis ratio, $B/A$, and ellipsoidal shapes via two axis ratios $p=b/a$ and $q=c/a$.
As $A$ and $a$ represent the longest axis, in both the 2D and 3D cases, axis ratios are constrained to $<1$ by definition.

Since observations of individual galaxies provide a single measure of the projected shape, 3D shape inferences may be obtained through statistical methods applied to galaxy samples \citep[e.g.][]{Binney1978EllipticalsProlateOblate, Vincent2005ShapeLuminosityProfile,Padilla2008ShapesInSloan}. 
The simplest methods rely on a comparison between the observed axis ratio distributions of a given galaxy sample and the expected distribution for projections of randomly oriented ellipsoids \citep[e.g.][]{Sandage1970IntrinsicFlatteningESOSpirals,Fasano1991IntrinsicShapesEllipticals,Kimm2007IntrinsicAxisRatios}. 
Using this method one can infer the 3D shape distribution of a given galaxy sample by varying the assumed underlying shape distribution of the randomly oriented ellipsoids until the resultant 2D shapes are well matched to the observations. 
There remains a significant degeneracy underlying such methods since a 2D shape distribution may be produced by a variety of underlying 3D shapes due to projection effects. Axisymmetry (i.e. $a=b$) may be assumed to break this degeneracy.

Resolved galaxy kinematics represent a hopeful avenue to break longstanding degeneracies \citep[e.g.][]{Binney1985TestingTriaxialityKinematics,Franx1991OrderedNatureEllipticals,Statler1994UncoveringIntrinsicShapes}. 
Early attempts at utilising kinematic maps in shape recovery relied on radio interferometry, a technique that is limited to very nearby galaxies \citep[e.g.][]{Bak2000IntrinsicShapeDistribution}. 
Detailed computationally intensive modelling of the kinematics and flux distribution is often required to constrain the intrinsic shape of individual galaxies \citep[e.g.][]{Statler1994UncoveringIntrinsicShapes, vandenBosch2009}. These usually require expert know-how and assumptions to break degeneracies.

With the more recent proliferation of integral field spectroscopic (IFS) instruments, resolved kinematics now exist for large samples of galaxies covering a broad range of galaxy properties \citep[e.g.][]{Cappellari2011Atlas3DIOverview,Croom2012SAMIOverview,Sanchez2012CALIFAOverview,Foster2021MAGPIOverview}.
Indeed, data from IFS surveys have been used to infer 3D shape distributions from a variety of samples \citep[e.g.][]{Weijmans2014ShapesOfEarlyTypes,Foster2017SAMIIntrinsicShapes,Li2018MaNGAIntrinsicShapes,Ene2018MASSIVEIntrinsicShapes}.

One major difficulty in testing the reliability of such efforts, however, is that the intrinsic shape distributions of observational samples is not known a priori. 
In this sense, the results of 3D shape recovery must place their faith in the dependability of the theoretical underpinning of the kinematics based methods when presenting the resulting 3D shape distributions.
One can, however, provide meaningful tests of the method by performing a series of mock observations of galaxies drawn from large scale, cosmological simulations from which the true, 3D shapes are already known. 
Such a test was performed on the methods of \citet{Franx1991OrderedNatureEllipticals} in \citet{Bassett2019GalaxyShapes}, who found that while $q$ can be recovered reasonably well for most galaxy types, $p$ is poorly recovered with the output distribution strongly favouring $p=1$ in all cases. 

The reason for the disproportionate recovery of $p=1$ in \citet{Bassett2019GalaxyShapes} was identified to be the assumed relationship between the intrinsic kinematic misalignment angle ($\psi_{int}$) and underlying galaxy shape \citep{Weijmans2014ShapesOfEarlyTypes}. 
To reduce the number of unknown parameters within the fitted model, it was assumed that $\psi_{int}$ was solely determined by the underlying 3D shape of the galaxy, something that was not borne out in the simulations, and thus need not be true in nature.
We are free from making such assumptions when using a non-linear, machine learning approach.
Using a variety of photometric and kinematic measurements commonly derived for IFS observations of galaxies, we can probe the shape information locked into various parameters, and quantify their shape-determining ``power'' relative to one another. 

In recent works, we have seen an increased uptake in the use of machine learning to explore astrophysical questions. 
For example, the primary drivers for individual cosmological parameters is being explored with the Cosmology and Astrophysics with MachinE Learning Simulations \citep[CAMELS;][]{VillaescusaNavarro+:2010}; for studies of hierarchical accretion histories and unlocking the un-observable merger history of a given galaxy \citep[e.g.,][]{Bottrell+:2022}; and for classifying morphology in large astronomical data sets in efficient ways \citep[e.g.,][]{Cavanagh+:2021}.
The machine learning approach is particularly useful when combined with large cosmological models of galaxy evolution, providing ground truth for labels that can then be applied to real data. 

In this work, we revisit the findings of \citet{Bassett2019GalaxyShapes} and attempt to use machine learning to recover the underlying 3D shape of individual galaxies without underlying assumptions.
As was done by \citet{Bassett2019GalaxyShapes}, we use galaxy simulations, in which the true underlying shapes of galaxies are known, and mock IFS observations of these objects, such that commonly used kinematic measurements can be derived. 
With these data, we train a neural network model, called a mixture density network (MDN), to recover the underlying axis ratios, and hence the 3D shape, of an observed system.

In this paper, we describe the construction of the training set in Section~\ref{sec:constructdata}, outline the development of the machine learning algorithm in Section~\ref{sec:pqml}, highlight our results in Section~\ref{sec:results}, discuss the ability of the algorithm to recover $p$ and $q$ and future directions in Section~\ref{sec:discussion}, and provide a summary in Section~\ref{sec:conclusion}.

\section{Construction of the Training Dataset} \label{sec:constructdata}

\begin{figure*}
\centering
\includegraphics[width=0.9\textwidth]{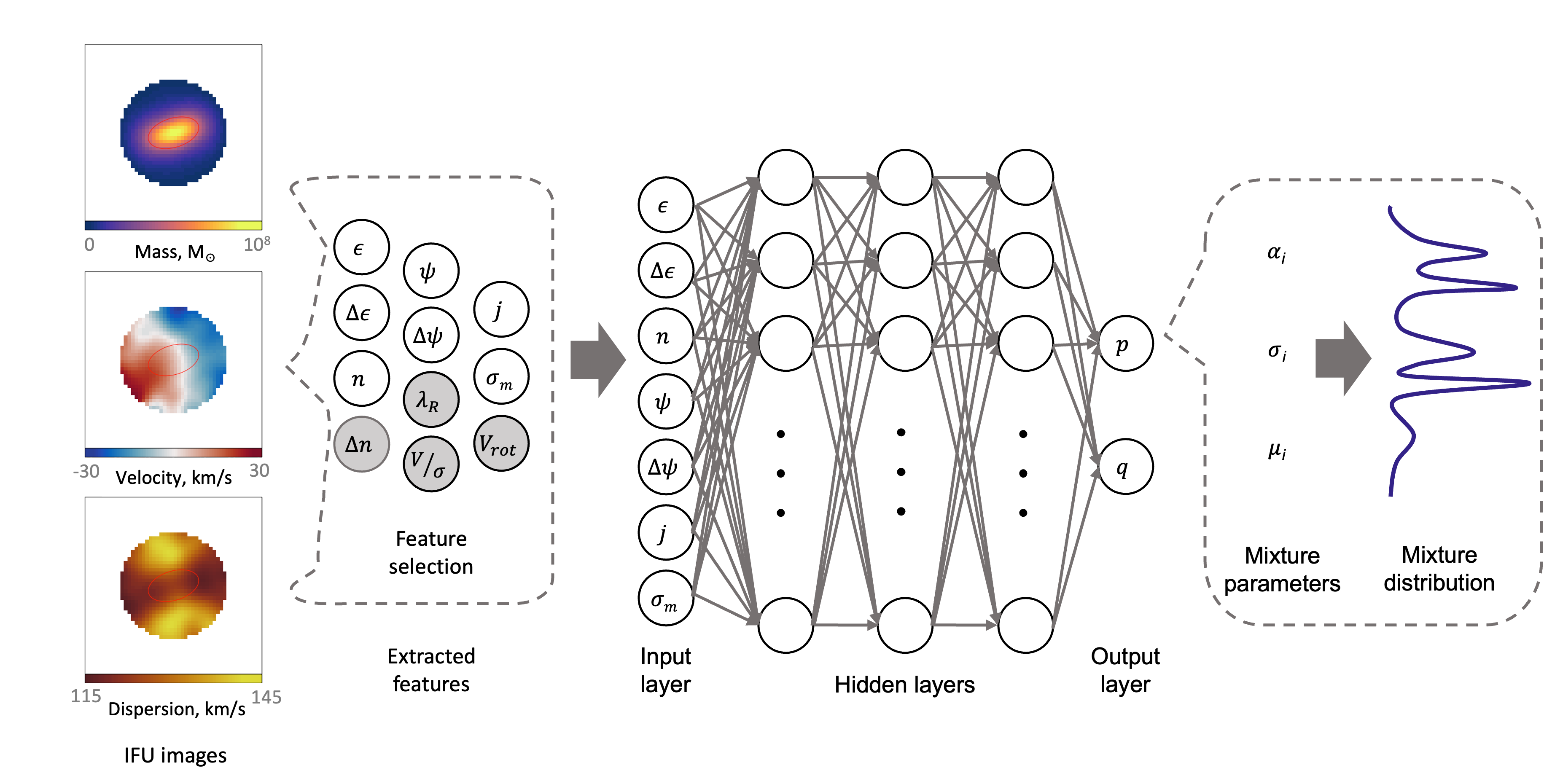}
\caption{Flowchart of the intrinsic shape determination pipeline. For each simulated galaxy, mock IFS images are constructed from which the kinematic and photometric features are extracted. Principal component analysis is applied for feature selection to choose a number of important features (those not selected are in grey). These are then fed to the mixture density network with 3 dense hidden layers of 128 nodes each. In the last layer, the MDN outputs a linear combination of Gaussian mixture parameters given by the weights $\alpha_{i}$, standard deviations $\sigma_{i}$, and means $\mu_{i}$ to predict the $p$ and $q$ distributions.}
\label{fig:pipelinemdn}
\end{figure*}

We perform post-processing of $z=0$ subhalo particle data from the \eagle{} cosmological, hydrodynamic simulation \citep{Schaye2015EAGLE, Crain2015EAGLE} to produce both 3D measurements and 2D mock-IFS observations.
Mock observations of this simulation are produced using the open-source code, \simspin{} \citep{Harborne2023SimSpin}, for 2519 galaxies extracted from \eagle{.} 
We measure common kinematic and photometric observables from these mocks, which are then used along with their ground truth 3D measurements, to train, validate and test the machine learning model.
A schematic diagram of our pipeline is illustrated in Figure~\ref{fig:pipelinemdn}.

Below, we begin by describing the methodology of computing the 3D measurements (Section~\ref{sec:3Dmeasures}) and constructing the mock observations (Section~\ref{sec:eagle}). 
We discuss how we have controlled for bias in our training data (Section~\ref{sec:unbias}) and how the individual kinematic and photometeric properties have been calculated for the training data set (Section~\ref{sec:trainingset}). 
We then outline the selection of parameters to be used for the MDN that have been chosen using a principal component analysis (PCA) technique (Section~\ref{sec:selectfeatures}).

\subsection{3D measurements} \label{sec:3Dmeasures}

We perform 3D shape measurements following the method described in \citet{Bassett2019GalaxyShapes}, which is based on \citet{Bak2000IntrinsicShapeDistribution}.
Under the assumption that galaxies are well described by a 3D ellipsoid, we determine the axes lengths of each galaxy by calculating the eigenvectors and eigenvalues of the reduced inertia tensor \citep[similar to e.g.][]{Allgood2006ShapeDarkMatterHaloes,Li2018TheShapesIllustris}. 
The reduced inertia tensor, $I$, is defined as:
\begin{equation}\label{eq:I}
	I_{ij} \equiv \sum_{n} \frac{x_{i,n}x_{j,n}}{\tilde{r}_{n}},
\end{equation}
where the tensor sum is performed for each combination of orthogonal axis directions, $ij$, and $\tilde{r}_{n}$ is the 3D, ellipsoidal radius measured for stellar particle $n$,
\begin{equation}
\tilde{r}_{n}=\sqrt{x_{n}^{2}+(y_{n}/p)^{2}+(z_{n}/q)^{2}},
\end{equation}
where the major, intermediate, and minor axes are aligned with the $x$, $y$, and $z$ directions, respectively. 
The axis ratios $p=b/a$ and $q=c/a$ are computed as the ratios of the corresponding eigenvalues.

In practice, we must exclude particles at large radii, which can skew measurements of $p$ and $q$. 
We therefore measure the shapes of galaxies only using particles within the ellipsoidal half mass radius, $\tilde{r}_{e}$, for consistency with previous works \citep[e.g.][]{Bassett2019GalaxyShapes} and for fair comparison with the radius at which the kinematic and photometric properties will be measured within the mock observations.

This requires us to measure the 3D shapes as an iterative process because the ellipsoidal half mass radius must be determined prior to measuring the eigenvalues of the reduced inertia tensor.
As an initial guess, we assume that each galaxy is spheroidal and measure the galaxy shape within the spherical half mass radius. 
We determine the new set of particles within the ellipsoidal half mass radius defined by this shape measurement. 
This process is repeated until the shape measurement has converged, giving the final values of $p$ and $q$ for each galaxy. 
The resulting range of galaxy shapes, as measured for the systems extracted from \eagle{,} are shown in Figure \ref{fig:EAGLE_shapes}.

\subsection{EAGLE mock observations} \label{sec:eagle}

The \eagle{} simulation suite is a set of hydrodynamical cosmological simulations designed to explore the necessary physical ingredients to form the galaxies we observe in the Universe today. 
A full description of these simulations can be found in \cite{Crain2015EAGLE} and \cite{Schaye2015EAGLE}. 
Individual galaxies have been extracted from snapshot 28 ($z = 0$) of the publicly available \texttt{RefL0100N1504} simulation box, a medium-resolution run of the ``reference'' model within a 100$^3$ comoving Mpc box. 
Using the \eagle{} database \citep{McAlpine2016EAGLEdatabase}, we have selected a sample of 3638 galaxies above a stellar mass limit of M$_{*} \geq 10^{10}$M$_{\odot}$. 
For each of these galaxies, we begin by measuring their shapes as described in Section \ref{sec:3Dmeasures}.

\begin{figure}
\centering
\includegraphics[width=\columnwidth]{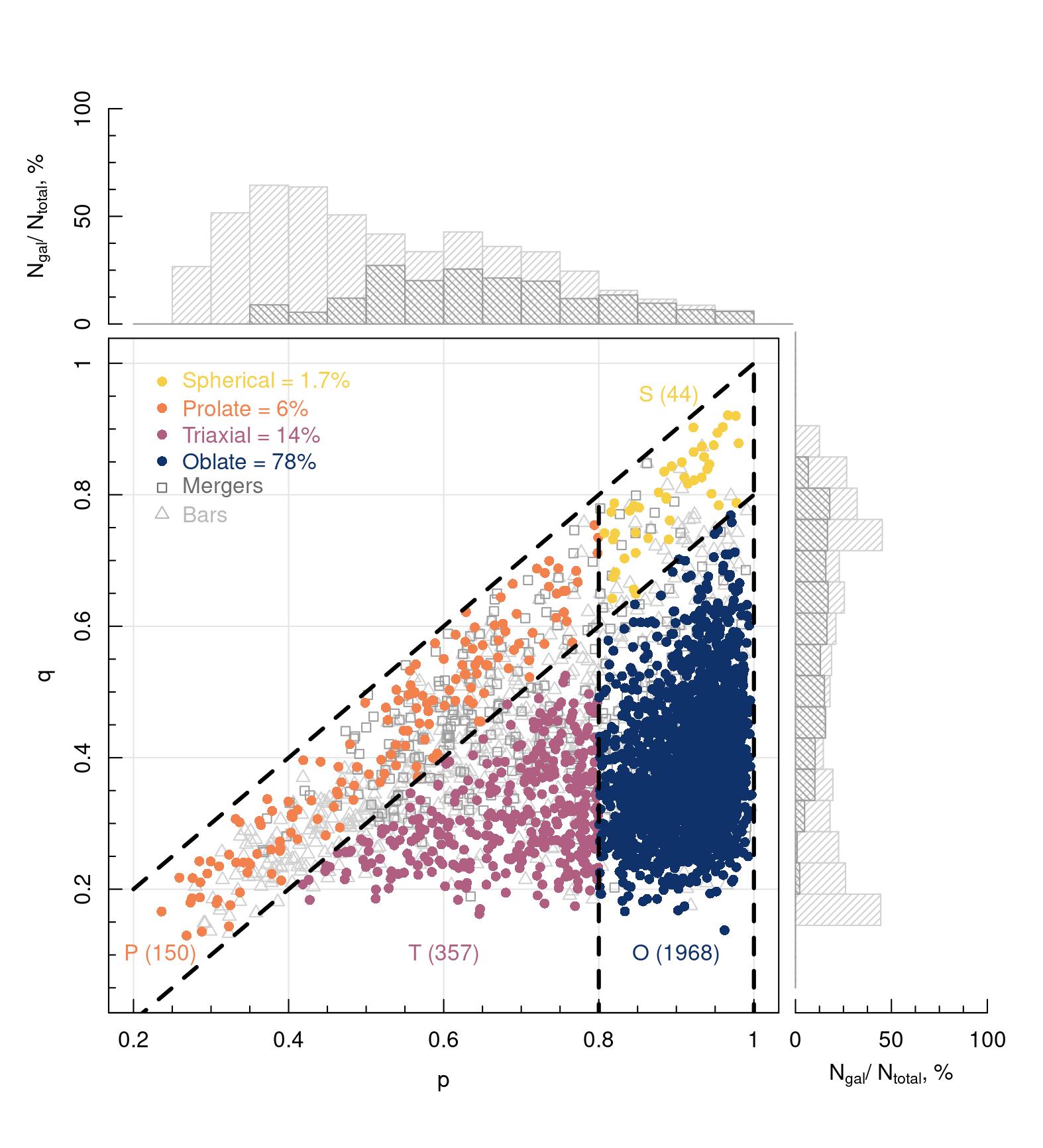}
\caption{The distribution of galaxy shapes measured from the \texttt{RefL0100N1504} box of the \eagle{} simulation. In dark grey squares we show galaxies that have undergone a major or minor merger within the last 5 Gyr, which we class as ``disturbed''. Light gray outline triangles show systems with significant bar structures. The histograms show the distribution of these barred and merger systems in $p$ and $q$ in light and dark grey respectively. Coloured circles represent galaxies that we have selected for our investigation. Each colour demonstrates the shape of the system, with spherical objects in yellow, prolate objects in orange, triaxial objects in pink and oblate objects in blue.}
\label{fig:EAGLE_shapes}
\end{figure}

The sample of galaxies is trimmed to remove any objects problematic for the algorithm design.
Galaxies marked as ``spurious'' or those which contain less than 50\% of the total stellar mass of the galaxy within a 50 kpc-radius sphere are also removed, leaving 3626 objects. 
Of these, we further restrict our sample to systems that have not experienced a major or minor merger (where we consider merger ratios $\geq$ 0.1) in the previous 5 Gyr of lookback time in order to avoid objects with significant disturbed features \citep{Morales2018SystematicSearchTidal}, leaving us with a sample of 3199 galaxies. 
As in \cite{Bassett2019GalaxyShapes}, we find that disturbed systems predominantly have low measured $q$ parameters and evenly distributed $p$ values, as shown in Figure \ref{fig:EAGLE_shapes}, and hence are more often interpreted as overly flat objects.
Finally, following the methodology of \cite{Bassett2019GalaxyShapes}, we remove any barred galaxies from the sample as the shapes of these systems within 1 \reff{} will often be mistaken for prolate structures. 
To find these systems, we measure the radial shape profiles for each galaxy and flag any systems whose shape changes from prolate or spherical to triaxial with increasing radius. 
Flagged galaxies are then visually inspected for any strong internal stellar features, such as bars or spiral arms.
This leaves us with a sample of 2519 galaxies.
The distributions of these barred and disturbed systems are demonstrated in the histograms in Figure \ref{fig:EAGLE_shapes}.

Mock IFS observations of the remaining \eagle{} galaxies are performed using the publicly available code \simspin{}  \footnote{\simspin{} is an open-source R-package at \url{https://github.com/kateharborne/SimSpin}}. 
While the methodology behind this code is outlined in \citealt{Harborne2023SimSpin}, we summarise the necessary details followed to construct each mock IFS observation here. 

We begin by initialising an observing telescope with properties equivalent to the \sami{} survey \citep{Croom2012SAMIOverview}. 
We have selected \sami{} for comparison with the work presented in \cite{Foster2017SAMIIntrinsicShapes} and \cite{Bassett2019GalaxyShapes}.
This is a predefined telescope type within the \simspin{} code. 
The spatial pixels of a \sami{}  observation are 0.5'' in width, with a equivalent velocity bin width of 1.04 \AA{} \citep{Green2017TheSAMISurvey}. 
The field-of-view of a \sami{} observation is circular with a diameter of 15''. 
For kinematics, observations from the blue arm of the AAOmega Spectrograph are used. 
The line spread function for this is well-approximated by a Gaussian of size 2.65 \AA{} measured at full-width at half-maximum \citep{vandeSande2017SAMIHigherOrderKinematics}.

With the telescope defined, each galaxy is placed at a specified distance from the observer.
A mass-weighted LOSVD is built for each spaxel dependent on the underlying velocity distribution of the particles at that pixel position. 
This information is then used to generate mock images of the projected mass, line-of-sight velocity and velocity dispersion in the style of a \sami{} observation. 
From these maps, we measure a series of parameters in order to train the model. 

To complement the kinematic maps, we also produce a series of mock $r$-band images using SimSpin.
This is done by associating stellar population templates with star particles of a given age and metallicity and computing the received flux through an $r$-band filter. 
These images are constructed with a spatial resolution equivalent to KiDS with 0.2''/pixel. 
Photometric properties, such as the observed half-light radii, ellipticity and \sersic{} index, are measured from each of these images. 

\subsection{Building the unbiased training set} \label{sec:unbias}

Using the defined sample of 2519 unique galaxies extracted from \eagle{,} we take a number of observations for each system from different orientation angles using this ``mock'' SAMI instrument in order to train and validate our algorithm. 
There are 13,265 independent observations, of which 85 percent are used in the training set and the remaining 15 percent are reserved for the validation set.
A further 9599 observations of \eagle{} galaxies are later used to test the algorithm. 
No unique galaxy appears in more than one of these sets (training, validation, or testing) in order to avoid confirmation bias. 

In each case, we wish to uniformly sample the $p-q$ parameter space so as not to bias our machine learning algorithm to one specific shape. 
If we provide a non-uniform distribution of galaxy shapes within the training set, the algorithm is likely to preferentially assign a new observation to an over-sampled parameter due to its prevalence in the training data, rather than due to its similarity with a specific shape's kinematic properties. 
Hence, we must ensure that the training set includes a uniformly distributed density of galaxies in the $p-q$ parameter space.

However, galaxy shapes are not equally prevalent in the Universe, or our simulations, as evident from Figure \ref{fig:EAGLE_shapes}. 
The majority of systems are oblate, while spherical systems are comparatively rare. 
In order to uniformly sample the parameter space, we have divided the $p-q$ space into 16 regions.  
We take repeat observations from different angles of galaxies in lower density $p-q$ regions in order to have a large enough data set to sufficiently train the algorithm.
No two observations of the same galaxy are from the same angle.

\begin{figure}
\centering
\includegraphics[width=\columnwidth]{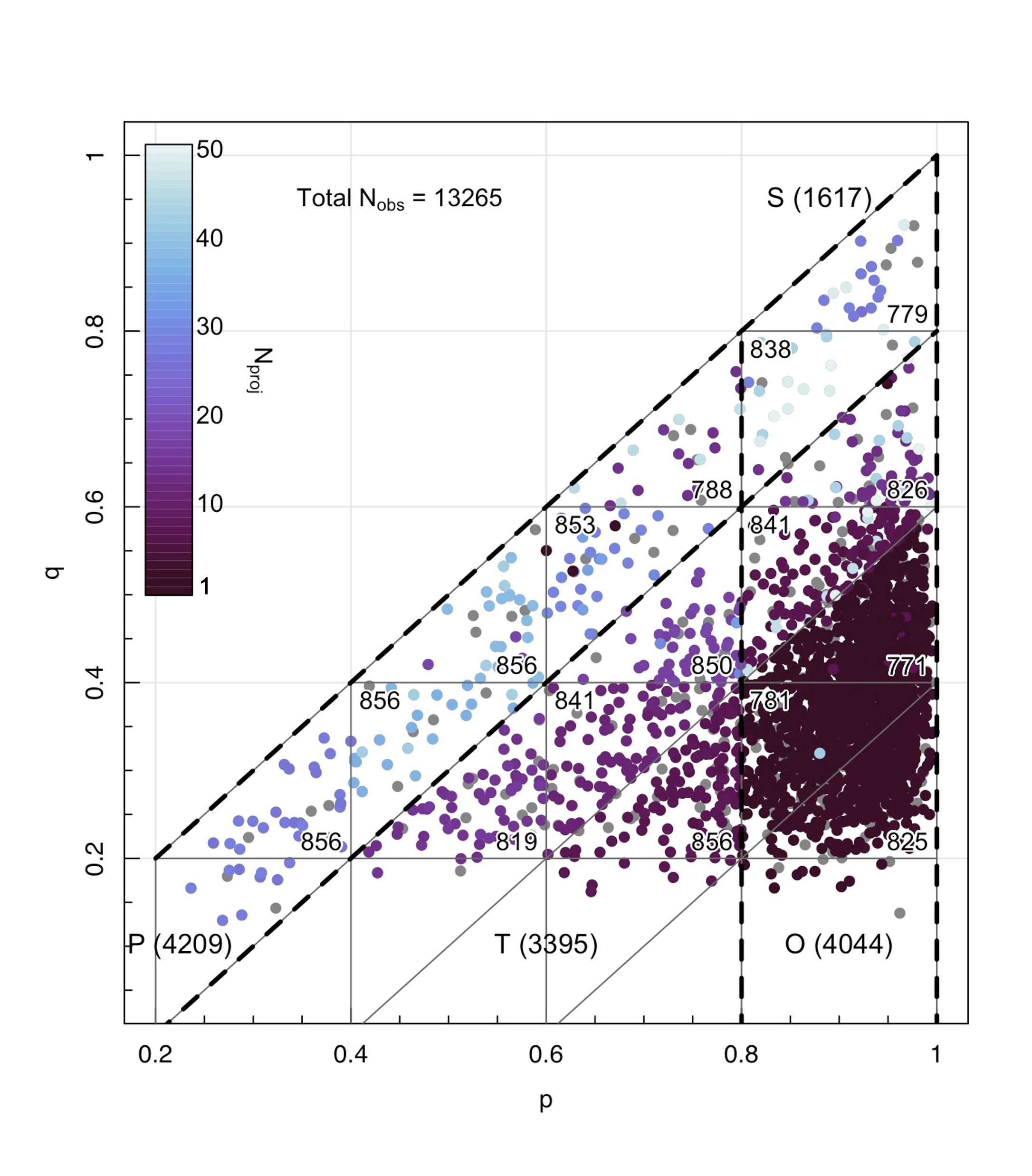}
\caption{Demonstrating the equal sampling of $p-q$ space within the training set. Each point shows an individual EAGLE galaxy within the full sample. Coloured points show galaxies selected for the training and validation sets, while grey points demonstrate galaxies that have been left for the testing phase. The colour of each point denotes the number of times that galaxy is observed in order to keep that $p-q$ region equally sampled. The number in the corner of each $p-q$ region demonstrates the total number of observations within that region.}
\label{fig:training_sample}
\end{figure}

In Figure \ref{fig:training_sample}, we demonstrate how we have sampled each region within $p-q$ space uniformly. 
It is noted that the $p-q$ space regions are equal sized triangles in the region $q > 0.4$ only, as in the lower $q < 0.2$ regime, particularly for prolate objects, we have significantly fewer available galaxies to sample. 
Objects that exist in the region $q <0.2$ have been combined with a neighbouring region above as demonstrated in Figure \ref{fig:training_sample}.

Further, we need to consider how the kinematic and photometric properties we measure for a given 3D shape is impacted by the chosen mock observational set up.
With a simulated system, we have the flexibility to re-orient the system in angle, projected distance and we can change the level of atmospheric seeing, in hope of recreating a reasonable range of mocks that reflect a real survey. 
We must take care that these choices do not introduce biases into the end result. 
In a follow up paper, we will demonstrate the importance of this unbiased sampling to produce a successful machine learning algorithm that uses kinematics from mock observations as input. 
For the sake of brevity, we present the optimal parameters over which to sample here. 

We create an unbiased kinematic training set by controlling "raw" observational parameters  including: 
(1) the orientation from which the galaxy is viewed, 
(2) spatial scales (adjusted by ``moving'' simulated galaxies closer and further away),  
and (3) the level of blurring caused by seeing conditions.  
These raw observation properties are cross-correlated with the observed kinematics and one another, such that it is necessary to perform multiple observations in a series of ordered steps to produce one unbiased element of training data.

It is important that these cross-correlated properties, the ``relative'' observational parameters, are fairly sampled.
Of course, within observational surveys, such properties cannot be controlled making it difficult to train such an algorithm with observations alone. 
In reality, the ground truth properties are also expensive and difficult to acquire.
For this reason, our simulations and \simspin{} are incredibly useful.
 
We select these parameters within reasonable limits based on the \sami{} survey. 
By ``reasonable limits'' we mean that we restrict our parameter space to values that would be considered reliable by the survey, rather than the full possible range of the instrument. 

\paragraph*{Projected orientation}
The observing angle for each run is randomly selected from the uniformly sampled surface of a sphere within each $p$-$q$ region.
No unique galaxy will be observed from the same observation angle more than once.
Within \simspin{,} this is controlled using two parameters: the \textit{inclination} about the x-axis of the projection, and the \textit{twist} about the y-axis of the projection.
We ensure that the combination of these parameters uniformly samples the surface of a sphere such that we are not biased towards a single observing angle in regions where more re-observations of single galaxies are necessary.
This also ensures that all possible 2D projections are evenly sampled within the training set.

\paragraph*{Spatial scales}
Physical spatial scales, i.e. the width of each spatial pixel in units of kpc, are varied by moving each galaxy to different distances. 
This distance is selected to sample a uniform distribution of pixels per measurement radius (otherwise called the ``spatial pixel scale'').
This is a ``relative'' observational property that we need to consider in relation to the projected angle above. 
If we were to require a uniform distribution of distances (the ``raw'' property) for each galaxy shape, we would inadvertently cause the distribution of spatial pixel scales to be lower for oblate systems (which, when viewed edge on have far fewer pixels within their measurement radius than an edge-on elliptical). 
The spatial pixel scale is known to correlate with the uncertainty of observed kinematic parameters such as the spin parameter $\lambda_R$ \citep[see Appendix C in ][]{Harborne2020Recoveringdata}.
We know that galaxy shape is correlated with the effects of seeing conditions on the observed kinematics, and the size of this effect in turn is dependent on the number of pixels within the measurement radius. 
We do not want to train our algorithm to learn that oblate systems are characterised solely in this way, else any observation with poor seeing and low spatial pixel sampling may be labelled ``oblate''. 
Hence, once a projected orientation is chosen for a galaxy, we adjust the projected distance to that object to recover a given spatial pixel scale selected from a uniform distribution. 

\begin{figure*}
\centering
\includegraphics[width=0.95\textwidth]{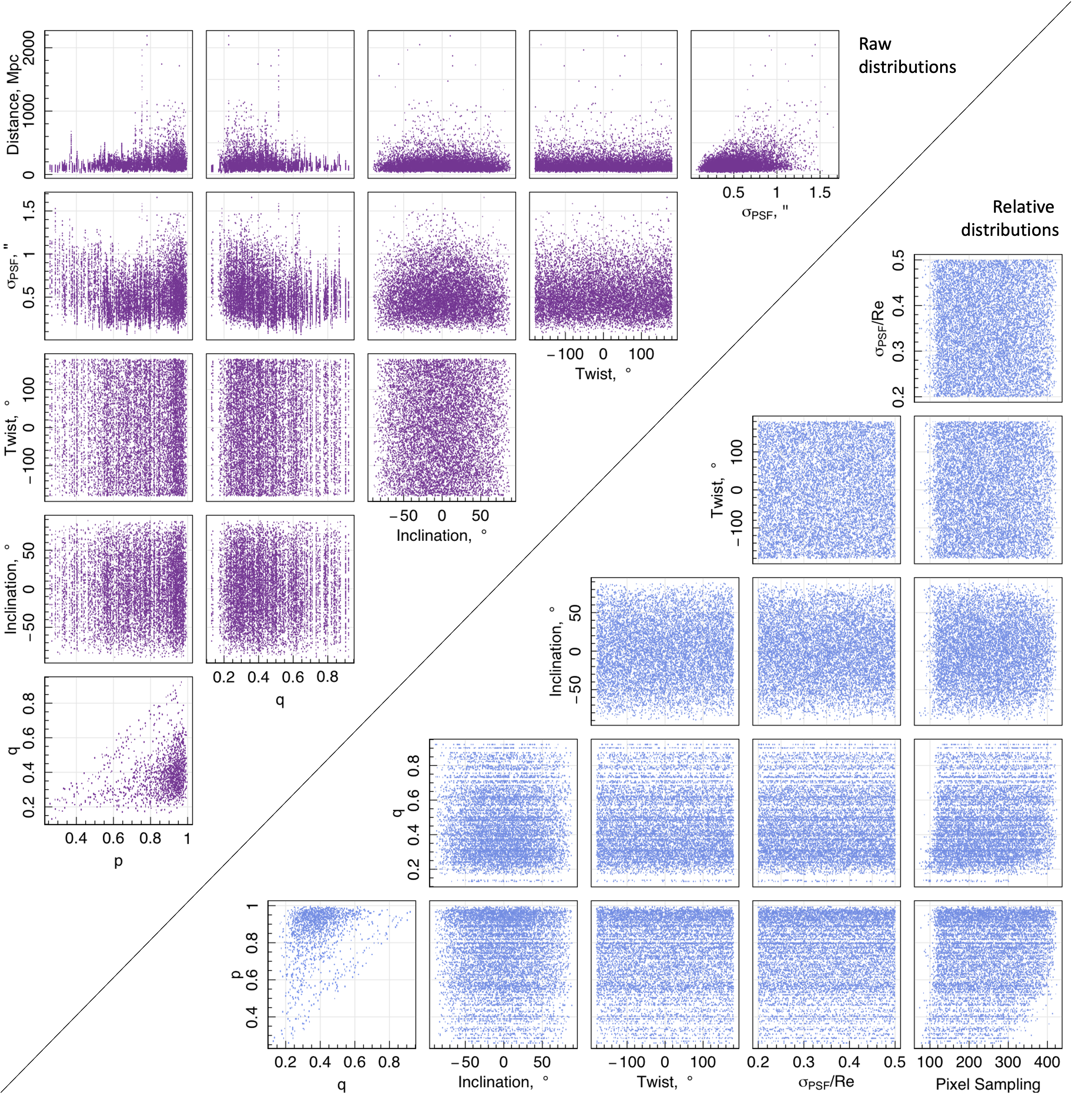}
\caption{(Top left) Considering the \textit{raw} distributions of tuneable observation properties controlled in each mock observation using a corner plot. The relationship between each property is demonstrated  in purple. By ``raw'' we mean the values modified per observation i.e. viewing angle, size of the PSF and projected distance to the object. (Bottom right) Considering the \textit{relative} distributions of the important tuneable observation properties to ensure mock observations are uniform in the important ratios, as shown in the blue corner plot. These ratios, i.e. the size of the PSF relative to the size of the object, and the number of pixels within the measurement radius, are important for measuring observable kinematics to produce an unbiased training set. The approximately uniform distribution shown in blue demonstrates that our sample selection is not biasing our algorithm results.} 
\label{fig:covariance}
\end{figure*}

In the lower right-hand corner of Figure \ref{fig:covariance}, there is a slight remaining trend with pixel sampling and very low $p$. 
This is due to the field of view limit of the SAMI observations. 
There is a limit to how large a flat, edge on disc can be projected within a circular aperture.
However, to remedy this, we have successfully flattened the distribution of pixel sampling with respect to the observational seeing conditions.
As a result, the effect on the observed kinematic parameters is minimised. 

\paragraph*{Seeing conditions}
Seeing conditions are uniformly sampled between 0.2 and 0.5 $\sigma_{PSF} / $R$_{maj}$, where we have chosen to define the size of the PSF relative to the measurement radius of the observation across a similar range to \sami{} observations.
Again, it has been shown in a number of works that the impact of seeing conditions on the observed kinematics of a galaxy is significant and varies with galaxy shape \citep{Harborne2020Recoveringdata, Graham2018SDSSproperties, Greene2018SDSSGalaxies, vandeSande2017SAMIHigherOrderKinematics, DEugenio2013FastLenticular}.

\vspace{12pt}

We use a Latin Hypercube Sampling (LHS)\footnote{We use the \texttt{lhs} R-package for this process, which can be found at \url{https://bertcarnell.github.io/lhs/index.html}.} procedure \citep{Stein1987LargeSampling, McKay1979ComparisonCode} to ensure that we do not introduce a correlation between these properties.
We demonstrate the success of this approach in the relative distributions on the right-hand side of Figure \ref{fig:covariance}.

Of course, while this careful selection produces a uniform distribution in the chosen ``relative'' parameters of importance to kinematic parameter recovery, this does result in skewed distributions for certain ``raw'' observed parameters, as shown by the distributions on the left-hand side in Figure \ref{fig:covariance}. 
We must put flatter objects with lower $p$ values closer to the observer (as shown by \textit{Distance, Mpc} in the final column). 
These closer objects see higher $\sigma_{PSF}$ values as a result. 

However, it is the relative relationship between $\sigma_{PSF}$ and the half-light radius of the galaxy that has impact on the measured kinematics and so it is this relative parameter that we must uniformly sample. 
Throughout the course of developing this MDN, several runs with different input training sets demonstrated how these biases propagate through the predictions of the network. 
This final set of relative distribution controls were found to best reflect reality of the shape distributions. 

\subsection{Computing the training parameters} \label{sec:trainingset}

\begin{figure*}
\centering
\includegraphics[width=0.95\textwidth]{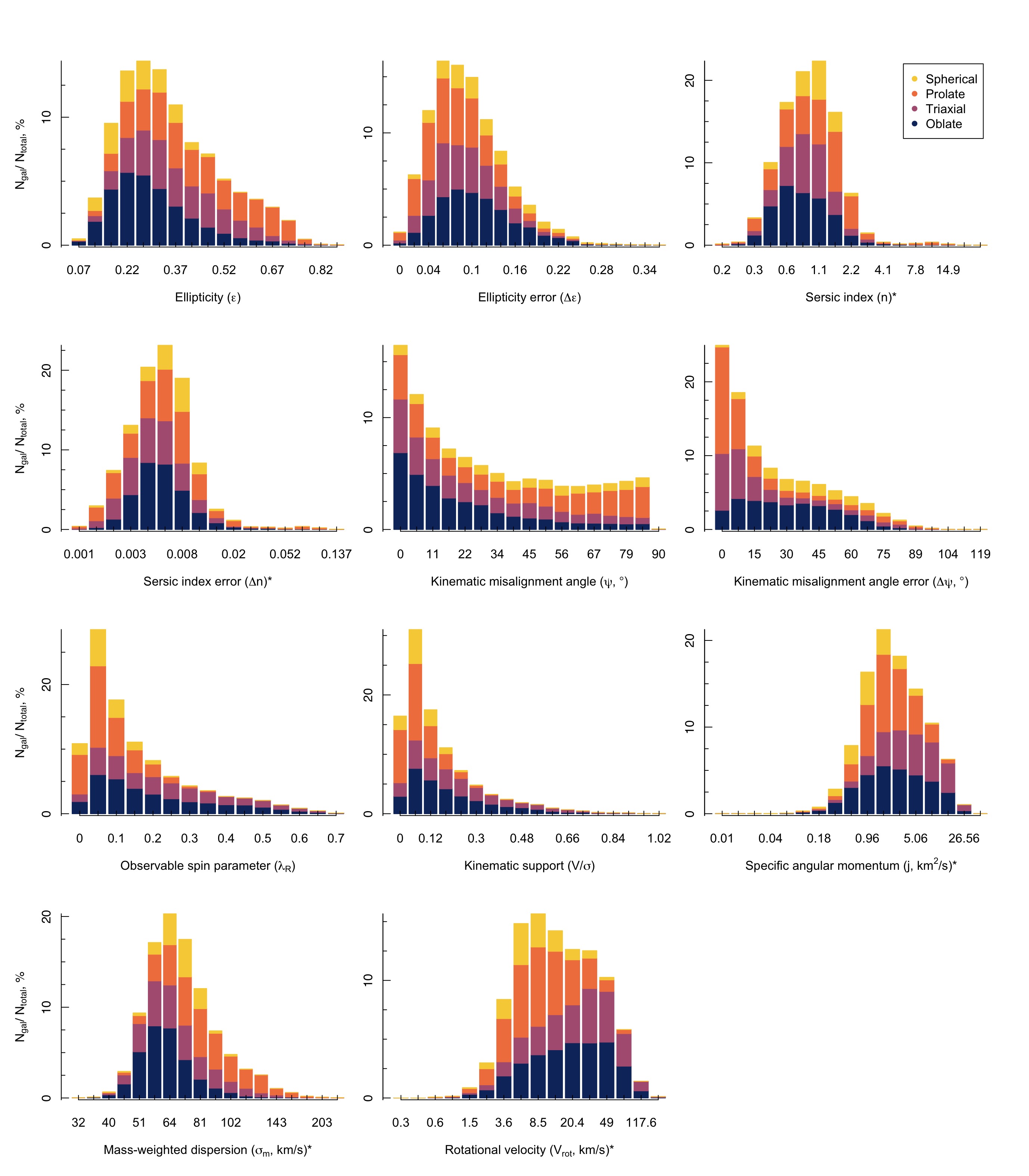}
\caption{Histograms showing how different intrinsic shapes of \eagle{} galaxies within our training data populate each observable parameter. In each case, the spherical systems are shown in yellow, prolate systems in orange, triaxial systems in purple and oblate systems in blue. The overall height of the bar shows the distribution of each kinematic parameter within the full training set. The coloured regions then demonstrate the percentage of each bar that is made up of each intrinstic shape. Starred (*) axis labels have been divided into equally-sized log10 bins to more clearly delineate between the groups, though bar labels are shown as the raw values for clarity. This plot demonstrates that, in none of the single measurements can we cleanly distinguish between the intrinsic shapes directly. This justifies the machine learning approach. }
\label{fig:parameters}
\end{figure*}

Using these unbiased mock IFS observations, we measure common kinematic and photometric parameters used in IFS surveys.
For each parameter, we demonstrate the ability of this term to distinguish unique 3D shapes using a histogram in Figure \ref{fig:parameters}.
The reason for selection and method of measuring each of these parameters is discussed below. 
While this is a comprehensive list of all the parameters explored, we note that the list of features used within the MDN model is trimmed down using principal component analysis. 
Only a selected number of key parameters are eventually used that best describe the features required to recover the intrinsic 3D shape distribution.
A list of the explored parameters as well as those selected for the mixture density network is summarised in Table~\ref{tab:features}.

\begin{table*}
\centering
\caption{List of investigated galaxy parameters in column 2 and their notation in column 1. A short description in column 3 and relevant sections where each parameter is described in the text are listed in column 4. Parameters that are selected as input features to the mixture density network (MDN) indicated in column 5.}
\label{tab:features}
\begin{tabular}{@{}cp{0.2\linewidth}p{0.4\linewidth}cc@{}}
\toprule
\textbf{Notation} & \textbf{Feature} & \textbf{Short Description} & \textbf{Section} & \textbf{Used in MDN} \\
(1) & (2) & (3) & (4) & (5)\\
\midrule
$\epsilon$ & Ellipticity & measured from each projected $r$-band luminosity map at the isophotal ellipse containing half the total flux of the galaxy & \S \ref{sec:ellipticity} & \checkmark \\ \midrule
$\Delta \epsilon$ & Ellipticity error & associated ellipticity error measured using the standard deviation between the ellipticities of the isophotal ellipses containing 40-60\% of the total galaxy flux & \S \ref{sec:ellipticity} & \checkmark \\ \midrule
$n$ & \sersic{} index & measured in the observed $r$-band light & \S \ref{sec:sersic} & \checkmark \\ \midrule
$\Delta n$ & \sersic{} index error & associated error on the fitted \sersic{} index where the error is computed as the standard deviation of the residuals to the fit & \S \ref{sec:sersic} & \checkmark \\ \midrule
$\psi$ & Kinematic misalignment angle & measured between the principle axis of rotation and the principle axis of the light distribution & \S \ref{sec:psi} & \checkmark \\ \midrule
$\Delta \psi$ & Kinematic misalignment angle error & associated error on the kinematic misalignment angle & \S \ref{sec:psi} & \checkmark \\ \midrule
$\lambda_{R}$ & Observable spin parameter & measured within 1 Re that describes the relative importance of rotation over dispersion as a function of radius in supporting a galaxy's structure & \S \ref{sec:lambdaR} & \\ \midrule
$V/\sigma$ & Kinematic support & measured within 1 Re of the relative importance of rotation over dispersion in supporting a galaxy's structure & \S \ref{sec:vsigma} & \\ \midrule
$j$ & Specific angular momentum & measured within 1 Re & \S \ref{sec:j} & \checkmark \\ \midrule
$\sigma_{m}$ & Mass-weighted average velocity dispersion & measured within 1 Re & \S \ref{sec:sigmam} & \checkmark \\ \midrule
$V_{\rm rot}$ & Observed rotational velocity & measured along the principle kinematic axis & \S \ref{sec:vrot} \\
\bottomrule
\end{tabular}%
\end{table*}


\subsubsection{Ellipticity, \texorpdfstring{$\epsilon$}{epsilon}}
\label{sec:ellipticity}
The ellipticity, $\epsilon$, of a galaxy in projection is the first parameter considered.
\citet{Franx1991OrderedNatureEllipticals} demonstrated how the observed ellipticity of a projected density distribution depends primarily on the intrinsic underlying galaxy shape and the viewing angle. 
For each intrinsic 3D shape, however, the mapping to a projected $\epsilon$ is not unique. 
This is seen in the first panel of Figure \ref{fig:parameters} where all 3D shapes may be projected to low apparent ellipticities ($\epsilon \leq 0.3$). 
Conversely, spherical systems cannot be projected to high ellipticities regardless of the inclination angle. 
This suggests that apparent ellipticity is likely to be a useful feature for certain regions of the $p$-$q$ parameter space.

To measure $\epsilon$, each mock KiDS $r$-band observation is fit using ProFound \citep{Robotham2018Profounddata}. 
We extract isophotal ellipses for each projection.
Photometric and kinematic measurements are computed within the average ellipse that contains 50\% of the total light of the galaxy, at which point the ellipticity of the object is also quoted. 
The ellipticity of this region is measured,
\begin{equation}
    \epsilon = 1 - (b/a),
\end{equation}
where the axis ratio ($b/a$) is computed by diagonalising the inertia tensor for all pixels within the half-light isophotal ellipse. 
The error on this parameter, $\Delta \epsilon$, is computed by taking the standard deviation of the ellipticity of concentric isophotes within 40-60\% of the total flux, as in \citet{Bassett2019GalaxyShapes}.
This is done for every inclination at which the galaxy is measured, hence why a disc may appear with a low ellipticity when viewed nearly face-on.

It is worth noting that we see very few highly elliptical objects from this sample of \eagle{} galaxies. 
This is not unexpected. 
As shown in \cite{Lagos2018Thegalaxies} and \cite{vandeSande2019SAMISimulations}, \eagle{} does not produce the observed high ellipticity objects we may expect. 
This is due in part to the temperature cooling floor imposed in cosmological hydrodynamical simulations that sets a minimum disc scale height of 1 kpc. 
There is also the effect of numerical heating \citep{Ludlow2019EnergySizes, Ludlow2021Spuriousparticles, Wilkinson2023impactdiscs}, which further increases the dispersion perpendicular to the plane of discs within the model, causing more disc-like objects to appear more elliptical in shape and kinematics. 
We can see that this is evident in the distribution of ellipticities. 
However, there is still a relative difference between each shape, so although the values may not reach as high as we may expect, their relative distribution seems reasonable. 

\subsubsection{\sersic{} index, \texorpdfstring{$n$}{n}}
\label{sec:sersic}
As a commonly measured structural parameter, we consider the \sersic{} index, $n$, in the set of training parameters. 
The \sersic{} index describes the slope of the light profile of a galaxy as fitted with a \sersic{} profile \citep{Sersic1963Influencegalaxy}. 
Discy systems correspond to low \sersic{} indices ($0.3 \leq n \leq 1.5$), while elliptical systems have higher values $n \sim 4$.
Despite its clear link to morphology and galaxy structure, Figure \ref{fig:parameters} suggests that the power of \sersic{} index in distinguishing 3D shapes is limited as we find the majority of shapes occurring in $0.3 \leq n \leq 4$. 
However, there is still a relative difference visible where spherical objects are seen to preferentially exist at the upper end of the limit and oblate objects dominate at the lower end of the limit, as we might expect.  

The \sersic{} profile is defined as:
\begin{equation}
\label{eq:sersic}
    I(R) = I_{e} \text{exp} \left\{ -b_{n} \left[ \left( \frac{R_m}{R_{e}} \right)^{1/n} - 1 \right] \right\}, 
\end{equation}
where the $I_{e}$ is the profile intensity measured at the half-light radius, $R_{e}$. 
The \sersic{} index, $n$, is a parameter that describes the shape of the light profile, while $b_{n}$ is used to ensure that, for a given \sersic{} index $n$, the correct integration properties occur at the half-light radius (i.e. that the function integrates to 0.5 for a given $n$).  
Finally, $R_{m}$ is the ``modified'' radius at which an isophote is defined. 
Within the ProFit code used to compute isophotal ellipses, this has a number of functional forms depending on whether the fitted isophotes are circular, elliptical or boxy.
For more details on each of the definitions, please refer to section 2.1.1 of \citet{Robotham2017ProFitimages}.
We use elliptical isophotes for this analysis. 

The associated error on $n$ is computed as the standard deviation of the residual between the fitted \sersic{} profile and the surface brightness profile. 

\subsubsection{Kinematic misalignment angle, \texorpdfstring{$\psi$}{psi}}
\label{sec:psi}
As suggested by a number of works \citep{Franx1991OrderedNatureEllipticals, Weijmans2014ShapesOfEarlyTypes, Foster2017SAMIIntrinsicShapes, Ene2018MASSIVEIntrinsicShapes, Bassett2019GalaxyShapes}, the kinematic misalignment angle, $\psi$, should also enable us to map back to the underlying 3D shape of the galaxy in question. 
Again, using $\psi$ alone is insufficient to invert back to an underlying 3D shape directly, as the kinematic misalignment is also dependent on the intrinsic axis of rotation.
For a triaxial galaxy, this rotation axis can lie anywhere in a plane that connects the longest and shortest axis of the system.
In the second row of Figure \ref{fig:parameters}, we consider the distribution of $\psi$ and its associated error with different intrinsic shape classifications for observations in our training set. 
It is clear that triaxial objects make up a similar percentage of every bin. 
Despite this, there are a number of clear trends with oblate and prolate objects predominantly occupying the low $\psi$ and high $\psi$ bins respectively, providing some level of distinguishing power. 
Using machine learning, we can explore the relative importance of this parameter with respect to other photometric and kinematic features. 

To compute $\psi$, we use the following equation from \citet{Franx1991OrderedNatureEllipticals}:
\begin{equation}
    \text{sin}( \psi ) = \left| \text{sin} \left( PA_{phot} - PA_{kin} \right) \right|.
\end{equation}

The photometric position angle, $PA_{phot}$, is measured as the orientation of the major axis of the isophotal ellipse in degrees from vertical. 
Similarly, the error on this value is computed using the standard deviation of isophotes' major axes angles for the isophotes 40-60\% of the total light. 

The kinematic position angle, $PA_{kin}$, describes the stellar angular rotation vector along which the rotational velocity is maximised. 
To compute $PA_{kin}$, we use the Kinemetry technique outlined in Appendix C of \citet{Krajnovic2006Galaxiesspectroscopic}. 

\subsubsection{Observable spin parameter, \texorpdfstring{$\lambda_{R}$}{lambdaR}}
\label{sec:lambdaR}

The observable spin parameter $\lambda_R$ was defined by \citeauthor{Emsellem2007Thegalaxies} in \citeyear{Emsellem2007Thegalaxies} to describe the level of coherent rotation in galaxies observed as part of the SAURON survey \citep{Bacon2001Thespectrograph}. 
$\lambda_R$ is now commonly measured by IFS survey teams as a method of distinguishing fast rotators from slow rotators within a $\lambda_R-\epsilon$ plane.
Given the underlying relation between the intrinsic shape and rotational support, as well as the commonality of this measure, it is an obvious parameter to include within the machine learning training. 
We can see strong evidence for this in the first panel of the third row of Figure \ref{fig:parameters}.
While there are relatively fewer high-spin objects observed in the \eagle{} training set, we can see that any objects above $\lambda_R$ of 0.3 are either oblate or triaxial systems. 

We use the following equation to define the spin parameter within the half-light isophotal ellipse defined in \S \ref{sec:ellipticity}:
\begin{equation}
    \label{eq:lr}
    \lambda_R = \frac{\sum_i^N F_i \; R_i \; |V_i|}{ \sum_i^N M_i \; R_i \sqrt{V_i^2 + \sigma_i^2}}, 
\end{equation}
where $F_i$ is the flux of stellar particles contained per pixel, $i$, $V_i$ is the LOS velocity, $\sigma_i$ is the LOS velocity dispersion and $R_i$ is the ellipsoidal radius corresponding to the semi-major axis of the ellipse at this $i$th pixel location. 
Only pixels contained within the half-light radius (described in \S \ref{sec:ellipticity}) are included within this calculation.
We note that this ellipsoidal radius definition is slightly different to the original $\lambda_R$ definition in \citet{Emsellem2007Thegalaxies}, which uses circularised radial weighting. The choice of an elliptical aperture is made to be consistent with the values measured by the SAMI survey \citep{vandeSande2017SAMISlowRotators}.  

\subsubsection{Kinematic support, \texorpdfstring{$V/\sigma$}{V/sigma}}
\label{sec:vsigma}
Along a similar vein, $V/\sigma$ is another common dynamical parameter used to gauge the balance of rotational vs. pressure support of galaxies \citep{Illingworth1977Rotationgalaxies, Binney1978EllipticalsProlateOblate, Davies1983Thegalaxies}.  
With the advent of IFS, this value was reformulated by \citet{Binney2005Rotationrevisited} using the tensor virial theorem in order to relate kinematics back to the intrinsic flattening.
For this reason, it is included within the training parameters. 

We use the light-weighted kinematics to compute $V/\sigma$ which is given by the equation:
\begin{equation}
\label{eq:vsigma}
    V/\sigma = \sqrt{\frac{\sum_i^N F_i \; V_i^2}{\sum_i^N M_i \; \sigma_i^2}}, 
\end{equation}
where, as before, $F_i$ is the stellar flux contained per pixel, $i$, $V_i$ is the LOS velocity at that pixel and $\sigma_i$ is the LOS velocity dispersion at that pixel. 
As with the measurement of $\lambda_R$, only pixels within the measurement radius of the half-light ellipse are used in this computation. 
When considering the distribution of this parameter across different galaxy shapes, in the central panel of the third row in Figure \ref{fig:parameters}, we find qualitatively similar distributions to the $\lambda_R$ parameter.

\subsubsection{Specific angular momentum, \texorpdfstring{$j$}{j}}
\label{sec:j}

In \citet{Bassett2019GalaxyShapes}, it was found that the intrinsic flattening of a galaxy was most strongly correlated with the stellar specific angular momentum, $j$. 
Considering the distribution of galaxy shapes in the right panel of the third row in Figure \ref{fig:parameters}, we can already see that high $j$ galaxies do exhibit flatter structures for our training galaxies taken from the \eagle{} simulation. 

We compute the specific angular momentum, $j$, using the formula:
\begin{equation}
\label{eq:j}
    j = \frac{\sum_i^N F_i \; R_i \; |V_i|}{\sum_i^N F_i},
\end{equation}
where the terms are the same as those used above. 
Again, only pixels contained within the half-light radius are included within this calculation.

\subsubsection{Mass-weighted velocity dispersion, \texorpdfstring{$\sigma_{m}$}{sigma}}
\label{sec:sigmam}
To then consider the level of dispersion support, in contrast to rotation and $j$, we compute the mass-weighted velocity dispersion, $\sigma_m$ within the half-light isophote. 
As can be seen in Figure \ref{fig:parameters}, the first panel in the final row shows increasing preponderance of prolate systems with increasing dispersion. 

This parameter is calculated using the LOS-velocity dispersion maps for each observation:
\begin{equation}
    \sigma_m = \frac{\sum_i^N M_i \; \sigma_i}{\sum^N_i M_i},
\end{equation}
where here, $\sigma_i$ is the LOS velocity dispersion and $M_i$ is the stellar mass per pixel, $i$, contained within the half-light radius. 

\subsubsection{Rotational velocity, \texorpdfstring{$V_{\rm rot}$}{Vrot}}
\label{sec:vrot}

The final flavour of kinematic parameter used is the rotational velocity, $V_{\rm rot}$.
This is another way to parameterise the level of rotation support for a system and was commonly recovered for long-slit spectroscopy and radio interferometry measurements \citep{Davies1988TheGalaxies, Bak2000IntrinsicShapeDistribution}. 
To compute this parameter for our galaxies, we use the kinematic position angle returned for the computation of $\psi$ and realign our galaxy velocity map such that the maximum gradient is horizontal within the image.
The central row of pixels from the velocity image is then used to visualise the galaxy rotation curve, with the left and right of the curve normalised and averaged.
$V_{\rm rot}$ is then quoted at the turnover of the rotation curve.

We can see from the final panel in Figure \ref{fig:parameters} that oblate and triaxial systems often occupy these higher $V_{\rm rot}$ regions.

\subsection{Selecting input features} \label{sec:selectfeatures}

\begin{figure}
\centering
\includegraphics[width=\columnwidth]{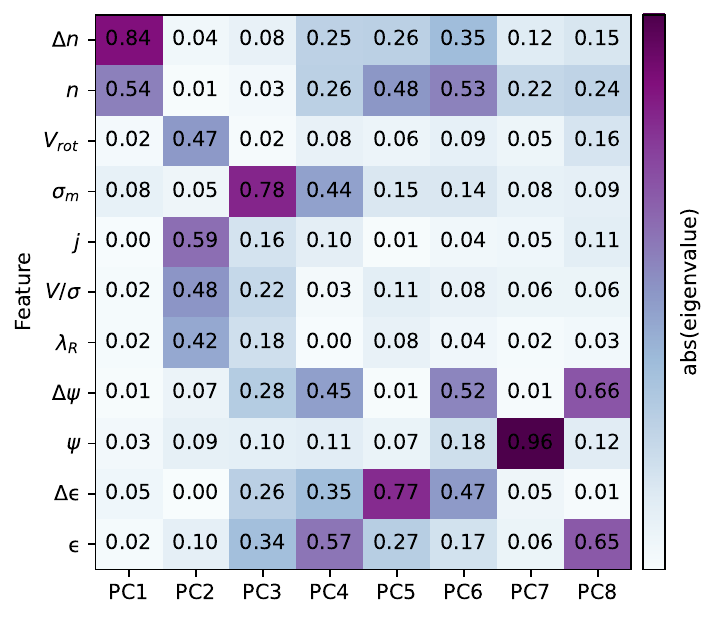}
\caption{Feature selection of galaxy kinematic parameters (see Table~\ref{tab:features} for notation) using principal component analysis. Absolute eigenvalues of the associated principal component (PC) are labelled and coloured, where a value of 1 indicates the strongest possible contribution with darker gradient.}
\label{fig:heatmap_featureimportancepca}
\end{figure}

From the initial set of 11 galaxy structural and kinematic parameters listed in Table~\ref{tab:features}, we further perform feature selection using principal component analysis \citep[PCA;][]{Pearson:1901} on the training data set.
Before applying PCA, all parameters of the training data are scaled based on the median and interquartile range between the 25th and 75th quantiles as a preprocessing step.
This PCA step is useful in determining which parameters hold the most constraining power and which sets may exhibit redundancies when tackling intrinsic shapes.
The resulting eigenvalues from the PCA quantifies the relative contributions of the various parameters to the main principal components (PCs).
The larger the absolute value of the corresponding eigenvalue, the more influence the parameter has on the PC.

Figure~\ref{fig:heatmap_featureimportancepca} shows the distribution of PCs along with weightings of each parameter on the PC and corresponding eigenvalues.
We select up to 8 PCs, which explain $\sim$99\% of the variance in the training data.
By performing this step, we can filter out redundant parameters and select those that are significant, which can lead to the overall improvement in the performance of the predictive model (see \ref{apd:predwofs} for comparative result without performing feature selection).
For each PC, the feature with the highest absolute eigenvalue is used as input feature to train the neural network model described in the next section.
This corresponds to 8 features ({\em Used in MDN} column in Table~\ref{tab:features}), namely ellipticity $\epsilon$, ellipticity error $\Delta \epsilon$, kinematic misalignment angle $\psi$, kinematic misalignment angle error $\Delta \psi$, specific angular momentum $j$, mass-weighted average velocity dispersion $\sigma_{m}$, \sersic{} index $n$, and \sersic{} index error $\Delta n$.

\section{Recover \texorpdfstring{$p-q$}{p-q} Distribution using Machine Learning} \label{sec:pqml}

\subsection{Building the mixture density network (MDN)}

In order to recover $p$ and $q$ as probability distributions, we build a mixture density network \citep[MDN;][]{Bishop:1994}
using \textsc{Keras} \citep{Chollet:2015}, an open-source \textsc{Python} package for deep learning, with \textsc{TensorFlow} \citep{Martin+:2015} backend.
MDN combines a deep or fully connected neural network with mixture of distributions.
It is a feed-forward neural network that maps a set of input features, $\vec{x}$, to generate output, $y$, of mixture models \citep{McLachlan+Basford:1988}.
Generally, the mixture distribution is represented by a Gaussian Mixture Model (GMM).
The conditional probability density function can be expressed as:
\begin{equation}
p(y|\vec{x})=\sum^{m}_{i=1} \alpha_{i}(\vec{x})\mathcal{N}(y;\mu_{i}(\vec{x}),\sigma^{2}_{i}(\vec{x})),
\label{eqn:mdncprobd}
\end{equation}
where $m$ is the number of mixture components, and $\alpha_{i}(\vec{x})$, $\mu_{i}(\vec{x})$ and $\sigma^{2}_{i}(\vec{x})$ are the corresponding weight, mean and variance of the $i$-th component of the mixture.

The network architecture consists of several layers, starting with an input layer, followed by dense hidden layers characterised by a non-linear activation function, and lastly an output layer with GMM, as illustrated in Figure~\ref{fig:pipelinemdn}.
For the hidden layer, we use three dense layer networks with 128 neurons and rectified linear unit as activation function.
Additionally, a dropout layer with 15\% drop rate is inserted between each hidden layer to reduce over-fitting.

In the MDN output layer, we find that changing the number of mixture components does not significantly affect the accuracy and set it to be 5.
The GMM parameters are derived from the network output vector, $\vec{z}$ as follows:
\begin{align}
\alpha_{i}(\vec{x}) &= \frac{\exp(z^{\alpha}_{i})}{\sum^{M}_{j=1}\exp(z^{\alpha}_{j})}, \label{eqn:mdnalphai} \\
\sigma_{i}(\vec{x}) &= \exp(z^{\sigma}_{i}), \label{eqn:mdnsigmai} \\
\mu_{i}(\vec{x}) &= z^{\mu}_{i}. \label{eqn:mdnmui}
\end{align}
The softmax activation function in Equation~\ref{eqn:mdnalphai} ensures that the weights are positive and sum to one.
An exponential activation is applied in Equation~\ref{eqn:mdnsigmai} such that the standard deviations are positive.
For the means in Equation~\ref{eqn:mdnmui}, a linear activation is used.
During the training process, the GMM parameters are optimised by minimising the negative logarithmic likelihood of Equation~\ref{eqn:mdncprobd} as the loss function using root mean squared propagation.
We train the MDN model for 5000 epochs using the training data set.
Early stopping is imposed to terminate the run when there is no significant improvement in the model performance's on the validation set.
Finally, the optimised GMM parameters are used to predict the distributions of $p$ and $q$ of the test dataset.

\subsection{Computing evaluation metrics}

To evaluate the performance of the model predictions, we compute the commonly used root mean squared error (RMSE) metric. The RMSE is the average of the squares of the difference between the actual value, $y$, and predicted value, $\hat{y}$, given by:
\begin{equation}
\mathrm{RMSE}=\sqrt{\frac{1}{N_{\mathrm{gal}}}\sum_{i=1}^{N_{\mathrm{gal}}} (\hat{y}_{i}-y_{i})^{2}},
\end{equation}
where $N_{\mathrm{gal}}$ is the number of galaxies in the dataset. The mean of the predicted $p$ and $q$ distributions are used to compare to the actual values of $p$ and $q$.

Additionally, we also assess the degree of success and contamination.
For each predicted shape class, we compute the negative predicted value (NPV) defined as:
\begin{equation}
\label{eqn:NPV}
    \text{NPV} = \frac{\text{Number of TRUE negatives}}{\text{Number of negative calls}}, 
\end{equation}
and the positive predicted value (PPV) or precision is defined as:
\begin{equation}
\label{eqn:PPV}
    \text{PPV} = \frac{\text{Number of TRUE positives}}{\text{Number of positive calls}}. 
\end{equation}
As such, the NPV describes the number of systems that have correctly been identified as NOT a given shape as a fraction of all the systems identified as NOT that shape, i.e. larger negative predicted rates imply a smaller contamination of this shape into other shape classes. 
By contrast, the PPV indicates the number of correctly identified objects of a given shape, i.e. the degree one can trust the shape given by the predicted $p$ and $q$ as a function of galaxy class.

\section{Results} \label{sec:results}

\begin{figure}
\centering
\includegraphics[width=0.95\linewidth,keepaspectratio]{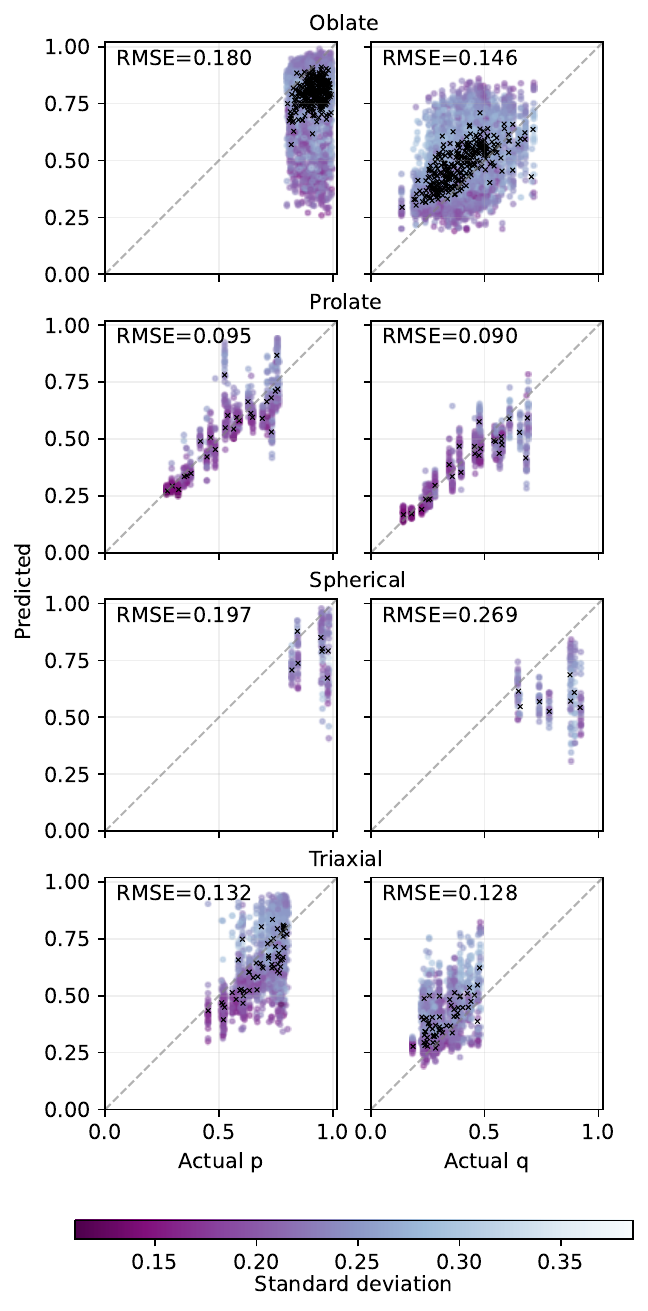}
\caption{Predicted against actual $p$ and $q$ for each galaxy shape using mixture density network (MDN) for the test data set. The black crosses represent the average prediction, while circles represent projections of individual galaxies colour coded by the standard deviation from the MDN output. The darker the gradient, the more certain. The prediction error is evaluated by the root mean squared error (RMSE), where lower values represent better agreement. For reference, the identity is shown as a grey dashed line. Although there is a large variation in the standard deviation of individual prediction, in most of the systems, the average of the predictions are close to the actual value.}
\label{fig:predvstrue_galclass}
\end{figure}

\begin{figure}
\centering
\includegraphics[width=\linewidth]{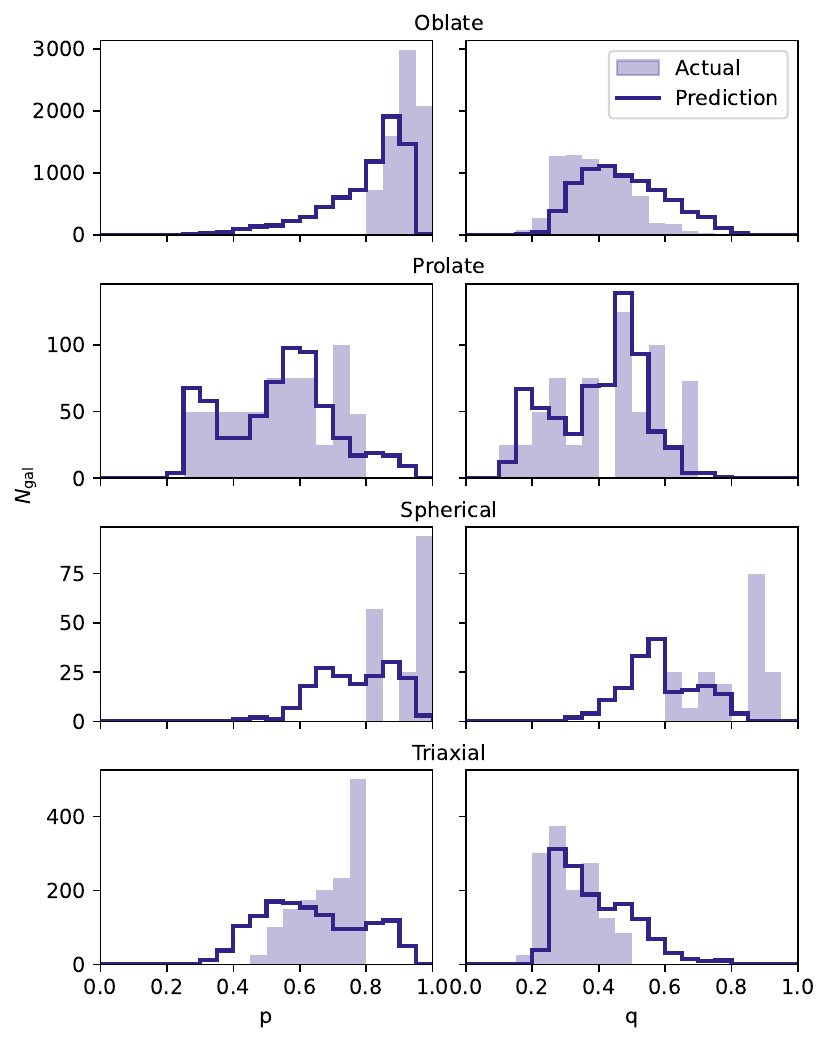}
\caption{Distributions of $p$ and $q$ recovered from the mixture density network model for each galaxy shape compared to the actual for the test data set.}
\label{fig:hist_galclass}
\end{figure}

Figure~\ref{fig:predvstrue_galclass} presents the predicted mean of $p$ and $q$ for each galaxy system from the MDN model in comparison to the actual values.
An indication of the uncertainty given by the standard deviation output from the Gaussian mixture parameter is shown in coloured points.
The corresponding distributions are shown in Figure~\ref{fig:hist_galclass}.

For oblate systems, which are the most abundant class objects, the predicted $p$ values tend to be underestimated, while the $q$ values tend to be overestimated.
The distribution of $p$ is highly skewed towards low values.
Prolate galaxies are well constrained with the recovered $p$ and $q$ distributions closely matching those of the actual distributions.
Spherical objects, on the other hand, are much harder to predict with both $p$ and $q$ predictions often lower than the actual values.
Their recovered distributions are also poorly estimated.
For triaxial objects, $p$ is better predicted than $q$, whereby the latter is overestimated in most cases.
It is also worth noting that there is a large scatter in the standard deviation, which can extend up to $\sim 0.37$.
However, the predicted means (indicated by the crosses in Figure~\ref{fig:predvstrue_galclass}) remain close to the true values.

Generally, the MDN is able to recover the underlying shape distributions of $p$ and $q$ for most of the systems, particularly prolate objects yield the best recovery with the lowest RMSE.
This is followed by triaxial, oblate, and lastly spherical objects with the largest RMSE.
Comparing the distributions between the actual and predicted for oblate and triaxial systems, even though their distributions are not exact, the MDN does manage to capture the overall shape of the distribution.
For example, both actual and predicted $p$ distributions for the oblate galaxies have similar negatively skewed shaped distribution.
The $q$ distributions for the triaxial galaxies appear to be positively skewed.

\begin{figure}
\centering
\includegraphics[width=\linewidth]{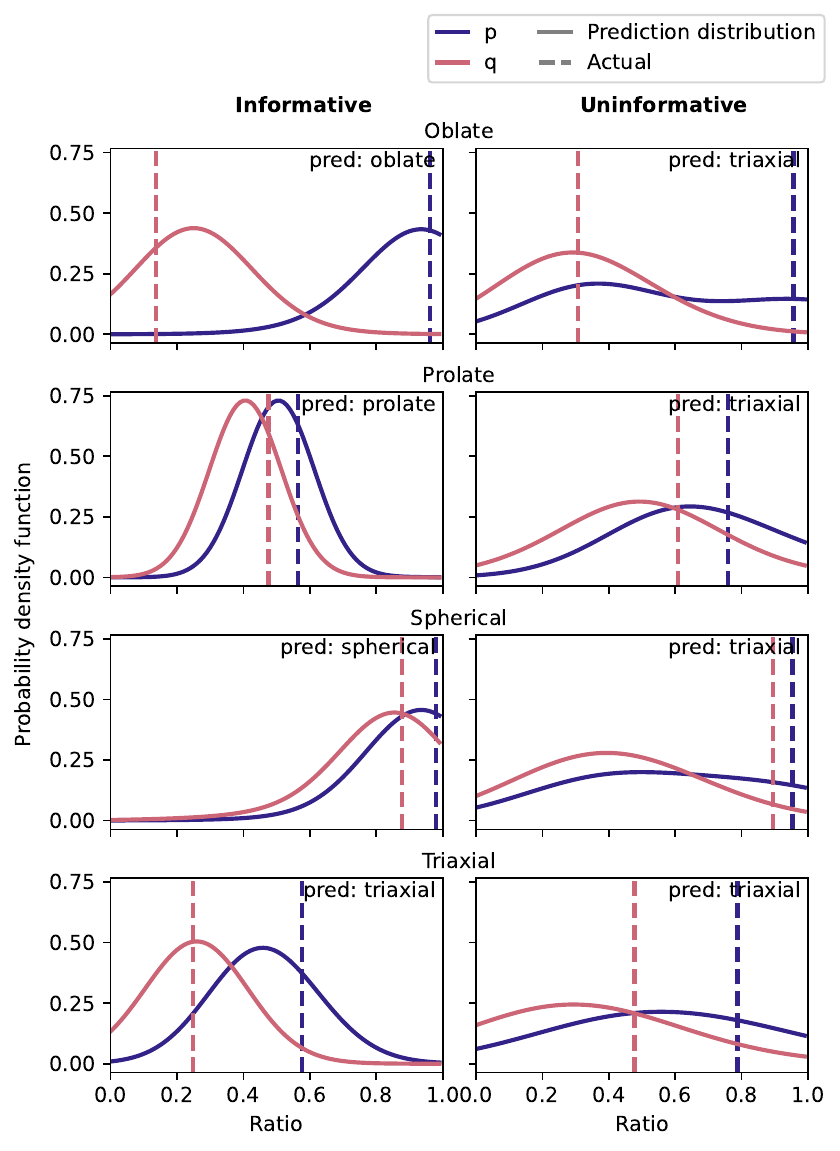}
\caption{The recovered $p-q$ shape probability density function within one standard deviation ($\sigma$) from the mixture density network, showing examples of ``informative'' (left column) and ``uninformative'' (right column) predictions for each galaxy shape. ``Informative'' fits are those with low standard deviation of $\sigma \leq 0.24$, while vice versa for ``uninformative'' fits. The predicted shape is shown on the top right of each panel. The vertical dashed line shows the actual value.}
\label{fig:pdftrue_galclass}
\end{figure}

For each prediction, the MDN outputs a probability density function that gives a useful indication of the allowable range of $p$ and $q$ values and whose standard deviations serve as estimates of the uncertainties.
Accounting for this, we further group them into ``informative'' and ``uninformative'' based on the $p-q$ errors.
Examples of the recovered shape distributions are portrayed in Figure~\ref{fig:pdftrue_galclass}.
For objects that are predicted with low uncertainties of less than the mean of the standard deviation at $\sigma \leq 0.24$, we labelled them as ``informative'', as shown in the left column.
The rest, that do not satisfy that criteria, are labelled as ``uninformative'' in the right column.
Generally, it can be seen that when the retrieved distribution of $p$ is broad (i.e. high standard deviation), the retrieved distribution of $q$ will also be broad, with Pearson correlation coefficient between the uncertainties of $p$ and $q$ of 0.796 ($p$-value $\ll 0.01$\%).
This suggests that both $p$ and $q$ predictions are less reliable if either one has large error.

\begin{figure}
\centering
\includegraphics[width=\columnwidth]{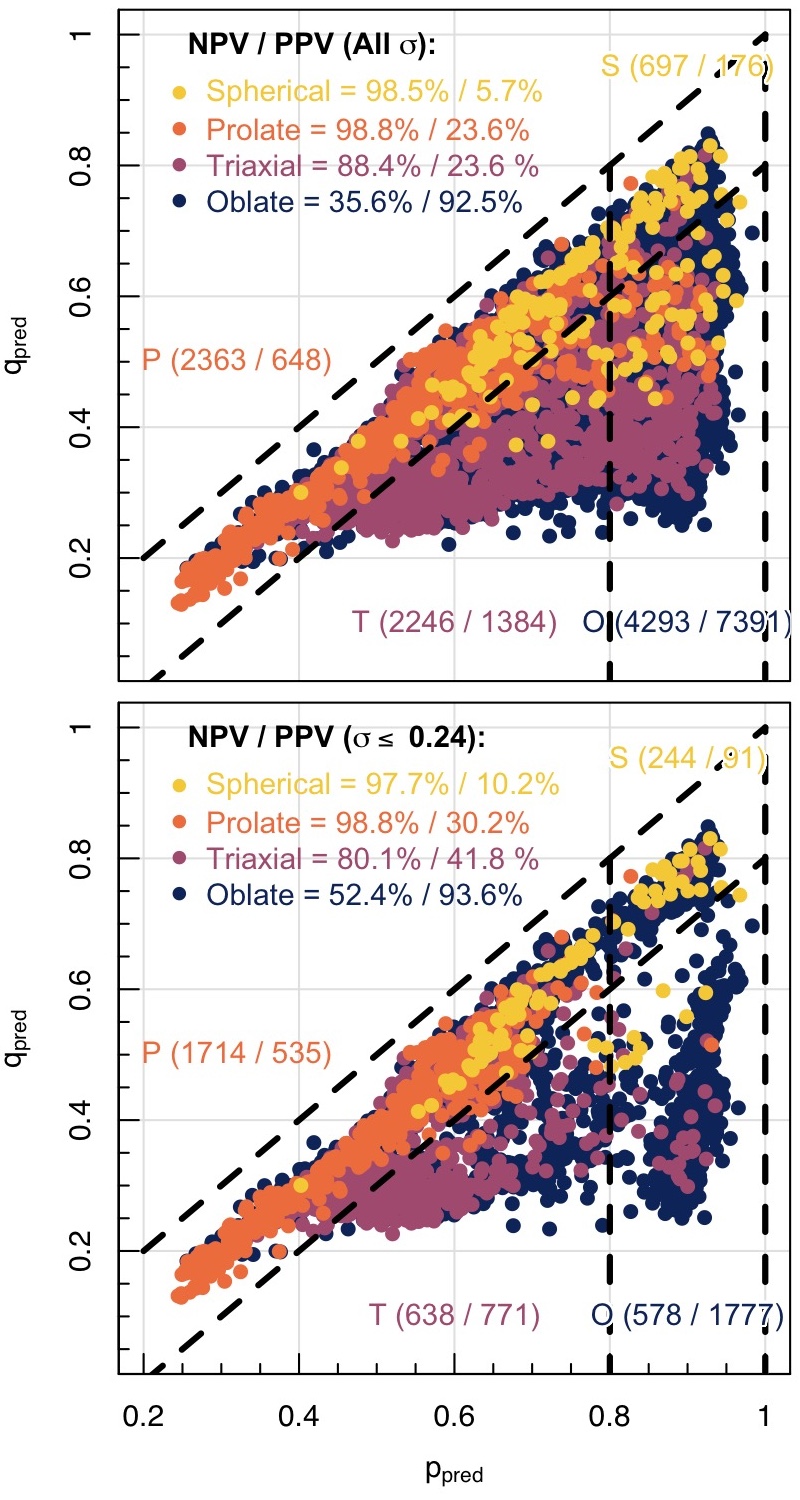}
\caption{Predicted $p$ and $q$ for each galaxy shape using mixture density network for all results (above) and for objects with a certainty of $\sigma \leq 0.24$ (below). Points are coloured by their true 3D shape class. The total number of predicted objects is indicated by the number in the round bracket in each shape region, followed by the number of true objects of that shape. The negative predicted values (NPV) and positive predicted values (PPV) are shown as a function of galaxy shape within the legend of each plot. We see the expected distributions broadly approaching the true as we select more certain predictions, though uncertain oblate systems (shown in blue) contaminate every other class in both plots. The reasons for this are discussed in \S \ref{sec:discussion}. }
\label{fig:predp+q_galclass}
\end{figure}

\begin{table}
\centering
\caption{The negative predicted values (NPV) and positive predicted values (PPV) for each underlying 3D shape given as percentages, where a higher value is better. In the left column of each metric, we show the values when all returned shapes are considered. In the right column of each metric, we consider only results with uncertainty less than 0.24.}
\label{tab:results_stats}
\begin{tabular}{@{}lcccc@{}}
\toprule
\textbf{Shape} & \textbf{NPV} (all) & \textbf{NPV} ($\sigma \leq 0.24$) & \textbf{PPV} (all) & \textbf{PPV} ($\sigma \leq 0.24$) \\
\midrule
Oblate & 35.6\% & 52.4\% & 92.5\% & 93.6\% \\
Prolate & 98.8\% & 98.8\% & 23.6\% & 30.2\% \\
Spherical & 98.5\% & 97.7\% & 5.7\% & 10.2\% \\
Triaxial & 88.4\% & 80.1\% & 23.6\% & 41.8\% \\
\bottomrule
\end{tabular}
\end{table}

In Figure~\ref{fig:predp+q_galclass}, we show the predicted $p$ and $q$ values of each observation in our testing set coloured by its true underlying 3D shape for all objects (top panel) and objects in the ``informative'' group (bottom panel).
Generally, the NPV and PPV improve as more certain (lower $\sigma$) predictions are filtered (see \ref{apd:trendsigma} for trends in NPV/PPV with $\sigma$).
Visually, we see that predicted classes are correlated with their true underlying shapes, with contamination specifically from oblate systems being predicted as every other shape class.
To quantify the degree of the success and contamination, we also compute the predictive values, namely NPV (Equation~\ref{eqn:NPV}) and PPV (Equation~\ref{eqn:PPV}), which are quoted as percentages in Table~\ref{tab:results_stats}.

Certain galaxy shape classes are harder to predict than the others.
Some classes, like spherical galaxies, are rarer and constitute a smaller proportion or relative area (see Figure~\ref{fig:EAGLE_shapes}) of the evenly sampled $p-q$ space, and this is also true in our Universe.
Based on the predictive values, we can be confident of the MDN predictions for oblate galaxies, which exhibit the highest PPV. 
We see that galaxies predicted to be oblate are very likely to truly be oblate, with $>92$\% of all systems in this region truly being this shape intrinsically. 
On the other hand, oblate systems have the lowest NPV rate, at $\sim 36$\%, demonstrating that an object classified as not oblate in shape has a high chance of actually being an oblate system. 
Though further improvement could be seen, to $\sim 52$\%, by selecting samples with higher certainty.
Despite contaminating the most of other regions in the predicted $p$-$q$ parameter space, very few other shapes are falling into this region.
By contrast, predominantly due to the contamination by oblate systems, we see the PPV values of all other shapes are significantly lower. 

At the other extreme, spherical systems have the lowest PPV at $<10$\% and thus lowest confidence.
This implies that many of the predicted spherical galaxies will be false positives.
Conversely, with a NPV of $>97$\%, we can confidently rule out a galaxy that is identified as not spherical to be true negative.
This is also the case for prolate and triaxial galaxies, which also present high NPV, but low PPV.
The prolate systems are predicted with particularly high NPV rate, at $\sim 99$\%, showing that an object identified as not prolate is very likely not to be intrinsically prolate.

\section{Discussion and Future Prospects} \label{sec:discussion}

As demonstrated in Figures \ref{fig:predvstrue_galclass}--\ref{fig:predp+q_galclass}, the machine learning approach to intrinsic shape recovery has a number of benefits over previous methods. 
Our approach returns a probability density function for every projection of individual galaxies.
This approach thus provides an estimate of the intrinsic shape along with uncertainties for individual galaxies.
Subsequently, this can be used to quantify our trust in the predictions.
We can be certain of our prediction for objects with accurately recovered $p$ and $q$ values with low standard deviation.
Those with large uncertainties are less reliable even for those with retrieved $p-q$ close to the actual.

In comparison to the original work of \citealt{Bassett2019GalaxyShapes}, we do not see a preferential recovery of $p=1$ in our predictions.
In particular, Figure \ref{fig:hist_galclass} in this work compares favourably to Figure 7 from \citealt{Bassett2019GalaxyShapes}. 
As suspected in that work, this improvement implies that $\psi$ is not uniquely linked to intrinsic shape - the addition of further kinematic and photometric information is necessary for realistic $p$-$q$ recovery.
Considering the results of the principal component analysis, we can see that $\psi$ is important to the determination, but in combination with a number of other parameters (Figure \ref{fig:heatmap_featureimportancepca}).
From Figure \ref{fig:parameters}, we also see that no single feature provides a clear mapping to the intrinsic 3D shape.

The kinematic misalignment $\psi$ does contribute to the absolute eigenvalues of the PCA, but only at the 7th component in Figure \ref{fig:heatmap_featureimportancepca}. 
Interestingly, we see that parameters such as the mass-weighted dispersion $\sigma_m$ and the specific angular momentum $j$ provide greater constraints to the intrinsic shape recovery than their combined counter-parts, $\lambda_R$ and $V/\sigma$.
We note that the dynamical range of each parameter has been controlled for through normalisation during the PCA, therefore differences in the parameter dynamical ranges cannot be driving these trends. 
This may be because a level of degeneracy is introduced by the division of parameters in $\lambda_R$ and $V/\sigma$ which may negate their shape describing power.

From Figure \ref{fig:predp+q_galclass}, we see that, despite clear improvements on previous approaches, we do struggle to recover some regions of the $p-q$ parameter space. 
This may in part be due to the formulation of the problem. 
Mathematically, we know that $p > q$, given $a > b > c$ (and $p = b/a$, $q = c/a$) and that both must be less than or equal to 1.
As the MDN returns a Gaussian distribution of predictions in $p$ and $q$, it is down-weighting the possibility of shapes that occur at these boundaries, especially the spherical objects where both ratios are near the boundary $p \sim q \sim 1$. 
This is clearly seen in Figure \ref{fig:predp+q_galclass}, where the predicted values fail to occupy any region close to the extremes of the parameter space.


The MDN uses a GMM, which may not be appropriate should the underlying distribution of shapes not be well characterised by Gaussians.
As can be observed in Figure \ref{fig:hist_galclass}, the expected distributions can vary between different systems. 
Oblate objects mainly have high $p$ values. However, the recovered distribution has a long tail towards lower values, which causes contamination to other galaxy classes in the $p-q$ space.
For spherical objects, the data appear sparse and modelling with multiple Gaussian components might not be a good representation of their distribution.
This has also been noted in \citealt{Bassett2019GalaxyShapes}, particularly for the $p-q$ recovery of spherical galaxies.
Additionally, based on our investigation, using a deeper neural network model and more mixture components in the MDN will not significantly improve the predictions for individual galaxy types.
A further exploration is to perform a detailed search of the MDN architecture and the best set and combinations of galaxy properties that can help in improving the machine learning model performance.
The addition of other kinematic and photometric properties, especially those known to correlate with structural parameters, such as colour, ionised gas kinematic parameters and star formation, might also provide additional information to aid in the recovery of $p$ and $q$.

This work presents a proof of concept that demonstrates the capability of MDN models to recover the $p$ and $q$ values to infer the shape of individual galaxy.
Instead of relying on restrictive assumptions about the relationship between the intrinsic shape and parameters such as the kinematic misalignment, $\psi$, we use the MDN model to simultaneously train on various kinematic and photometric parameters that provide meaningful properties to characterise 3D galaxy shapes.
Generally, the MDN model can recover the shape distribution for prolate objects, but performs more poorly for other shapes.
The MDN can provide estimates of the $p$ and $q$ values as well as meaningful uncertainties through probability distributions of those parameters.
We recommend the use of informative retrieved $p$ and $q$ values with low standard deviation, while cautioning against using those with large standard deviation.

The MDN pipeline requires low computational power to run, especially once it has been trained and large volumes of test data can be evaluated almost instantly.
Future work will include deploying MDN on mock data sets from a varied set of hydrodynamical cosmological simulations, such as \illustristng.
This would provide important insights on how differently the model might perform on real data, how well it scales across different data sets and also inform how different cosmological simulations compare with one another and with real galaxies.

\section{Conclusion} \label{sec:conclusion}

Recovering the underlying 3D intrinsic shapes of individual galaxies is a challenging task since our observations of galaxies appear as 2D projection maps.
As opposed to traditional methods that assume a relationship between the intrinsic kinematic misalignment and morphological axes, we employ a supervised machine learning method.
For this purpose, we build a mock observational dataset from the \eagle{} hydrodynamical cosmological simulation as our training and testing datasets.
We then apply a mixture density network (MDN) to predict the intrinsic axis ratios, $p$ and $q$, that characterise the 3D galaxy shape.

As with other methods, our approach also finds that it is harder to recover the distribution of $p$ than that of $q$, which can be retrieved reasonably well for the majority of systems.
We demonstrate that there is no simple pair of parameters that is able to provide a unique link to the 3D intrinsic shape.
We find instead that the inclusion and combination of both kinematic and photometric information does further improve the $p-q$ recovery.
We show that our MDN model has great potential in recovering the intrinsic shapes of the galaxies, along with subsequent uncertainties.
Our MDN approach can be applied to any integral field spectroscopic data. These works will serve as the foundation for future investigations on the use of ML for galaxy intrinsic shape recovery.

\begin{acknowledgement}
We thank the anonymous referee for valuable suggestions on the manuscript.

This project grew out of an Astronomy Data and Computing Services (ADACS) internship scholarship awarded to SY.
ADACS is funded from the Astronomy National Collaborative Research Infrastructure Strategy (NCRIS) allocation provided by the Australian Government and managed by Astronomy Australia Limited (AAL).
CF is the recipient of an Australian Research Council Future Fellowship (project number FT210100168) funded by the Australian Government.
JTM and CF are the recipients of ARC Discovery Project DP210101945.
Part of this research was conducted by the Australian Research Council Centre of Excellence for All Sky Astrophysics in 3 Dimensions (ASTRO 3D), through project number CE170100013.

This research was undertaken in part using the Setonix machine at the Pawsey Supercomputing Centre in Perth, Australia.
We also acknowledge the Virgo Consortium for making their simulation data available. 
The \eagle{} simulations were performed using the DiRAC-2 facility at Durham, managed by the ICC, and the PRACE facility Curie based in France at TGCC, CEA, Bruy\`{e}resle-Ch\^{a}tel. 
\end{acknowledgement}

\paragraph{Data Availability Statement}
The \eagle{} simulation is publicly available at \url{http://icc.dur.ac.uk/Eagle/}.
The code to generate 2D mock integral field spectroscopic observations is publicly available at \url{https://github.com/kateharborne/SimSpin}.
The codes used to perform the analysis presented in this work are available from the corresponding author upon reasonable request.

\printendnotes

\bibliography{galaxyshape_references}

\begin{thebibliography}{}
\expandafter\ifx\csname natexlab\endcsname\relax\def\natexlab#1{#1}\fi

\bibitem[{Abadi {et~al.}(2015)Abadi, Agarwal, Barham, Brevdo, Chen, Citro, Corrado, Davis, Dean, Devin, Ghemawat, Goodfellow, Harp, Irving, Isard, Jia, Jozefowicz, Kaiser, Kudlur, Levenberg, Man\'{e}, Monga, Moore, Murray, Olah, Schuster, Shlens, Steiner, Sutskever, Talwar, Tucker, Vanhoucke, Vasudevan, Vi\'{e}gas, Vinyals, Warden, Wattenberg, Wicke, Yu, \& Zheng}]{Martin+:2015}
Abadi, M., Agarwal, A., Barham, P., {et~al.} 2015, {TensorFlow}: Large-Scale Machine Learning on Heterogeneous Systems, software available from tensorflow.org

\bibitem[{Allgood {et~al.}(2006)Allgood, Flores, Primack, Kravtsov, Wechsler, Faltenbacher, \& Bullock}]{Allgood2006ShapeDarkMatterHaloes}
Allgood, B., Flores, R.~A., Primack, J.~R., {et~al.} 2006, MNRAS, 367, 1781

\bibitem[{Bacon {et~al.}(2001)Bacon, Copin, Monnet, Miller, Allington-Smith, Bureau, Carollo, Davies, Emsellem, Kuntschner, Peletier, Verolme, \& Zeeuw}]{Bacon2001Thespectrograph}
Bacon, R., Copin, Y., Monnet, G., {et~al.} 2001, Monthly Notices of the Royal Astronomical Society, 326, 23

\bibitem[{Bak \& Statler(2000)}]{Bak2000IntrinsicShapeDistribution}
Bak, J., \& Statler, T.~S. 2000, AJ, 120, 110

\bibitem[{Bassett \& Foster(2019)}]{Bassett2019GalaxyShapes}
Bassett, R., \& Foster, C. 2019, Monthly Notices of the Royal Astronomical Society, 487, 2354

\bibitem[{{Binggeli}(1980)}]{Binggeli1980}
{Binggeli}, B. 1980, \aap, 82, 289

\bibitem[{Binney(1978)}]{Binney1978EllipticalsProlateOblate}
Binney, J. 1978, Comments Astrophysics, 8, 27

\bibitem[{Binney(1985)}]{Binney1985TestingTriaxialityKinematics}
---. 1985, MNRAS, 212, 767

\bibitem[{{Binney}(2005)}]{Binney2005Rotationrevisited}
{Binney}, J. 2005, \mnras, 363, 937

\bibitem[{Bishop(1994)}]{Bishop:1994}
Bishop, C.~M. 1994

\bibitem[{{Bottrell} {et~al.}(2022){Bottrell}, {Hani}, {Teimoorinia}, {Patton}, \& {Ellison}}]{Bottrell+:2022}
{Bottrell}, C., {Hani}, M.~H., {Teimoorinia}, H., {Patton}, D.~R., \& {Ellison}, S.~L. 2022, \mnras, 511, 100

\bibitem[{Cappellari {et~al.}(2011)Cappellari, Emsellem, Krajnovi{\'c}, McDermid, Scott, Kleijn, Young, Alatalo, Bacon, Blitz, Bois, Bournaud, Bureau, Davies, Davis, de~Zeeuw, Duc, Khochfar, Kuntschner, Lablanche, Morganti, Naab, Oosterloo, Sarzi, Serra, \& Weijmans}]{Cappellari2011Atlas3DIOverview}
Cappellari, M., Emsellem, E., Krajnovi{\'c}, D., {et~al.} 2011, Monthly Notices of the Royal Astronomical Society, 413, 813

\bibitem[{{Cavanagh} {et~al.}(2021){Cavanagh}, {Bekki}, \& {Groves}}]{Cavanagh+:2021}
{Cavanagh}, M.~K., {Bekki}, K., \& {Groves}, B.~A. 2021, \mnras, 506, 659

\bibitem[{Chollet {et~al.}(2015)}]{Chollet:2015}
Chollet, F., {et~al.} 2015, Keras, \url{https://keras.io}

\bibitem[{{Contopoulos}(1956)}]{Contopoulos1956}
{Contopoulos}, G. 1956, \apj, 124, 643

\bibitem[{{Crain} {et~al.}(2015){Crain}, {Schaye}, {Bower}, {Furlong}, {Schaller}, {Theuns}, {Dalla Vecchia}, {Frenk}, {McCarthy}, {Helly}, {Jenkins}, {Rosas-Guevara}, {White}, \& {Trayford}}]{Crain2015EAGLE}
{Crain}, R.~A., {Schaye}, J., {Bower}, R.~G., {et~al.} 2015, \mnras, 450, 1937

\bibitem[{Croom {et~al.}(2012)Croom, Lawrence, Bland-Hawthorn, Bryant, Fogarty, Richards, Goodwin, Farrell, Miziarski, Heald, Jones, Lee, Colless, Brough, Hopkins, Bauer, Birchall, Ellis, Horton, Leon-Saval, Lewis, L{\'o}pez-S{\'a}nchez, Min, Trinh, \& Trowland}]{Croom2012SAMIOverview}
Croom, S.~M., Lawrence, J.~S., Bland-Hawthorn, J., {et~al.} 2012, Monthly Notices of the Royal Astronomical Society, 421, 872

\bibitem[{{Davies} \& {Birkinshaw}(1988)}]{Davies1988TheGalaxies}
{Davies}, R.~L., \& {Birkinshaw}, M. 1988, \apjs, 68, 409

\bibitem[{{Davies} {et~al.}(1983){Davies}, {Efstathiou}, {Fall}, {Illingworth}, \& {Schechter}}]{Davies1983Thegalaxies}
{Davies}, R.~L., {Efstathiou}, G., {Fall}, S.~M., {Illingworth}, G., \& {Schechter}, P.~L. 1983, \apj, 266, 41

\bibitem[{D'Eugenio {et~al.}(2013)D'Eugenio, Houghton, Davies, \& Bont{\`a}}]{DEugenio2013FastLenticular}
D'Eugenio, F., Houghton, R. C.~W., Davies, R.~L., \& Bont{\`a}, E.~D. 2013, Monthly Notices of the Royal Astronomical Society, 429, 1258

\bibitem[{Emsellem {et~al.}(2007)Emsellem, Cappellari, Krajnovi{\'c}, Ven, Bacon, Bureau, Davies, Zeeuw, Falc{\'o}n-Barroso, Kuntschner, McDermid, Peletier, \& Sarzi}]{Emsellem2007Thegalaxies}
Emsellem, E., Cappellari, M., Krajnovi{\'c}, D., {et~al.} 2007, Monthly Notices of the Royal Astronomical Society, 379, 401

\bibitem[{Ene {et~al.}(2018)Ene, Ma, Veale, Greene, Thomas, Blakeslee, Foster, Walsh, Ito, \& Goulding}]{Ene2018MASSIVEIntrinsicShapes}
Ene, I., Ma, C.-P., Veale, M., {et~al.} 2018, MNRAS, 479, 2810

\bibitem[{Fasano(1991)}]{Fasano1991IntrinsicShapesEllipticals}
Fasano, G. 1991, MNRAS, 249, 208

\bibitem[{Foster {et~al.}(2016)Foster, Pastorello, Roediger, Brodie, Forbes, Kartha, Pota, Romanowsky, Spitler, Strader, Usher, \& Arnold}]{Foster2016SluggsStellarKinematics}
Foster, C., Pastorello, N., Roediger, J., {et~al.} 2016, MNRAS, 457, 147

\bibitem[{Foster {et~al.}(2017)Foster, van~de Sande, D'Eugenio, Cortese, McDermid, Bland-Hawthorn, Brough, Bryant, Croom, Goodwin, Konstantopoulos, Lawrence, L{\'o}pez-S{\'a}nchez, Medling, Owers, Richards, Scott, Taranu, Tonini, \& Zafar}]{Foster2017SAMIIntrinsicShapes}
Foster, C., van~de Sande, J., D'Eugenio, F., {et~al.} 2017, MNRAS, 472, 966

\bibitem[{Foster {et~al.}(2021)Foster, Mendel, Lagos, Wisnioski, Yuan, D'Eugenio, Barone, Harborne, Vaughan, Schulze, Remus, Gupta, Collacchioni, Khim, Taylor, Bassett, Croom, McDermid, Poci, Battisti, Bland-Hawthorn, Bellstedt, Colless, Davies, Derkenne, Driver, Ferr{\'e}-Mateu, Fisher, Gjergo, Johnston, Khalid, Kobayashi, Oh, Peng, Robotham, Sharda, Sweet, Taylor, Tran, Trayford, van~de Sande, Yi, \& Zanisi}]{Foster2021MAGPIOverview}
Foster, C., Mendel, J.~T., Lagos, C. D.~P., {et~al.} 2021, PASA, 38, e031

\bibitem[{Franx {et~al.}(1991)Franx, Illingworth, \& de~Zeeuw}]{Franx1991OrderedNatureEllipticals}
Franx, M., Illingworth, G., \& de~Zeeuw, T. 1991, ApJ, 383, 112

\bibitem[{Graham {et~al.}(2018)Graham, Cappellari, Li, Mao, Bershady, Bizyaev, Brinkmann, Brownstein, Bundy, Drory, Law, Pan, Thomas, Wake, Weijmans, Westfall, \& Yan}]{Graham2018SDSSproperties}
Graham, M.~T., Cappellari, M., Li, H., {et~al.} 2018, Monthly Notices of the Royal Astronomical Society, 477, 4711

\bibitem[{Green {et~al.}(2017)Green, Croom, Scott, Cortese, Medling, D'Eugenio, Bryant, Bland-Hawthorn, Allen, Sharp, Ho, Groves, Drinkwater, Mannering, Harischandra, van~de Sande, Thomas, O'Toole, McDermid, Vuong, Sealey, Bauer, Brough, Catinella, Cecil, Colless, Couch, Driver, Federrath, Foster, Goodwin, Hampton, Hopkins, Jones, Konstantopoulos, Lawrence, Leon-Saval, Liske, L{\'o}pez-S{\'a}nchez, Lorente, Mould, Obreschkow, Owers, Richards, Robotham, Schaefer, Sweet, Taranu, Tescari, Tonini, \& Zafar}]{Green2017TheSAMISurvey}
Green, A.~W., Croom, S.~M., Scott, N., {et~al.} 2017, Monthly Notices of the Royal Astronomical Society, 475, 716

\bibitem[{Greene {et~al.}(2018)Greene, Leauthaud, Emsellem, Ge, Arag{\'o}n-Salamanca, Greco, Lin, Mao, Masters, Merrifield, More, Okabe, Schneider, Thomas, Wake, Pan, Bizyaev, Oravetz, Simmons, Yan, \& Bosch}]{Greene2018SDSSGalaxies}
Greene, J.~E., Leauthaud, A., Emsellem, E., {et~al.} 2018, The Astrophysical Journal, 852, 36

\bibitem[{Harborne {et~al.}(2020)Harborne, Sande, Cortese, Power, Robotham, Lagos, \& Croom}]{Harborne2020Recoveringdata}
Harborne, K., Sande, J. v.~d., Cortese, L., {et~al.} 2020, Monthly Notices of the Royal Astronomical Society, 497, 2018

\bibitem[{{Harborne} {et~al.}(2023){Harborne}, {Serene}, {Davies}, {Derkenne}, {Vaughan}, {Burdon}, {Lagos}, {McDermid}, {O'Toole}, {Power}, {Robotham}, {Santucci}, \& {Tobar}}]{Harborne2023SimSpin}
{Harborne}, K.~E., {Serene}, A., {Davies}, E.~J.~A., {et~al.} 2023, \pasa, 40, e048

\bibitem[{Hubble(1926)}]{Hubble1926}
Hubble, E.~P. 1926, The Astrophysical Journal, 64, 321

\bibitem[{{Illingworth}(1977)}]{Illingworth1977Rotationgalaxies}
{Illingworth}, G. 1977, \apjl, 218, L43

\bibitem[{Kimm \& Yi(2007)}]{Kimm2007IntrinsicAxisRatios}
Kimm, T., \& Yi, S.~K. 2007, ApJ, 670, 1048

\bibitem[{Krajnovic {et~al.}(2006)Krajnovic, Cappellari, Zeeuw, \& Copin}]{Krajnovic2006Galaxiesspectroscopic}
Krajnovic, D., Cappellari, M., Zeeuw, P. T.~d., \& Copin, Y. 2006, Monthly Notices of the Royal Astronomical Society, 366, 787

\bibitem[{Lagos {et~al.}(2018)Lagos, Schaye, Bahé, Sande, Kay, Barnes, Davis, \& Vecchia}]{Lagos2018Thegalaxies}
Lagos, C. d.~P., Schaye, J., Bahé, Y., {et~al.} 2018, Monthly Notices of the Royal Astronomical Society, 476, 4327

\bibitem[{{Lambas} {et~al.}(1992){Lambas}, {Maddox}, \& {Loveday}}]{Lambas1992}
{Lambas}, D.~G., {Maddox}, S.~J., \& {Loveday}, J. 1992, \mnras, 258, 404

\bibitem[{Li {et~al.}(2018{\natexlab{a}})Li, Mao, Cappellari, Graham, Emsellem, \& Long}]{Li2018MaNGAIntrinsicShapes}
Li, H., Mao, S., Cappellari, M., {et~al.} 2018{\natexlab{a}}, ApJL, 863, L19

\bibitem[{Li {et~al.}(2018{\natexlab{b}})Li, Mao, Emsellem, Xu, Springel, \& Krajnovi{\'c}}]{Li2018TheShapesIllustris}
Li, H., Mao, S., Emsellem, E., {et~al.} 2018{\natexlab{b}}, Monthly Notices of the Royal Astronomical Society, 473, 1489

\bibitem[{Ludlow {et~al.}(2021)Ludlow, Fall, Schaye, \& Obreschkow}]{Ludlow2021Spuriousparticles}
Ludlow, A.~D., Fall, S.~M., Schaye, J., \& Obreschkow, D. 2021, Monthly Notices of the Royal Astronomical Society, 508, 5114

\bibitem[{Ludlow {et~al.}(2019)Ludlow, Schaye, Schaller, \& Richings}]{Ludlow2019EnergySizes}
Ludlow, A.~D., Schaye, J., Schaller, M., \& Richings, J. 2019, Monthly Notices of the Royal Astronomical Society, 488, L123

\bibitem[{{McAlpine} {et~al.}(2016){McAlpine}, {Helly}, {Schaller}, {Trayford}, {Qu}, {Furlong}, {Bower}, {Crain}, {Schaye}, {Theuns}, {Dalla Vecchia}, {Frenk}, {McCarthy}, {Jenkins}, {Rosas-Guevara}, {White}, {Baes}, {Camps}, \& {Lemson}}]{McAlpine2016EAGLEdatabase}
{McAlpine}, S., {Helly}, J.~C., {Schaller}, M., {et~al.} 2016, Astronomy and Computing, 15, 72

\bibitem[{McKay {et~al.}(1979)McKay, Beckman, \& Conover}]{McKay1979ComparisonCode}
McKay, M.~D., Beckman, R.~J., \& Conover, W.~J. 1979, Technometrics, 21, 239

\bibitem[{{McLachlan} \& {Basford}(1988)}]{McLachlan+Basford:1988}
{McLachlan}, G.~J., \& {Basford}, K.~E. 1988, {Mixture models. Inference and applications to clustering} (M. Dekker New York)

\bibitem[{Morales {et~al.}(2018)Morales, Mart{\'\i}nez-Delgado, Grebel, Cooper, Javanmardi, \& Miskolczi}]{Morales2018SystematicSearchTidal}
Morales, G., Mart{\'\i}nez-Delgado, D., Grebel, E.~K., {et~al.} 2018, \aap, 614, A143

\bibitem[{Padilla \& Strauss(2008)}]{Padilla2008ShapesInSloan}
Padilla, N.~D., \& Strauss, M.~A. 2008, Monthly Notices of the Royal Astronomical Society, doi:10.1111/j.1365-2966.2008.13480.x

\bibitem[{Pearson(1901)}]{Pearson:1901}
Pearson, K. 1901, The London, Edinburgh, and Dublin Philosophical Magazine and Journal of Science, 2, 559

\bibitem[{Robotham {et~al.}(2018)Robotham, Davies, Driver, Koushan, Taranu, Casura, \& Liske}]{Robotham2018Profounddata}
Robotham, A.~S., Davies, L.~J., Driver, S.~P., {et~al.} 2018, Monthly Notices of the Royal Astronomical Society, 476, 3137

\bibitem[{Robotham {et~al.}(2017)Robotham, Taranu, Tobar, Moffett, \& Driver}]{Robotham2017ProFitimages}
Robotham, A. S.~G., Taranu, D.~S., Tobar, R., Moffett, A., \& Driver, S.~P. 2017, Monthly Notices of the Royal Astronomical Society, 466, 1513

\bibitem[{Ryden(2006)}]{Ryden2006IntrinsicShapeAtlas}
Ryden, B.~S. 2006, ApJ, 641, 773

\bibitem[{Sanchez {et~al.}(2012)Sanchez, Kennicutt, de~Paz, van~de Ven, V{\'\i}lchez, Wisotzki, Walcher, Mast, Aguerri, Albiol-Perez, Alonso-Herrero, Alves, Bakos, Bartakova, Bland-Hawthorn, Boselli, Bomans, Castillo-Morales, Cortijo-Ferrero, de~Lorenzo-Caceres, del Olmo, Dettmar, D{\'\i}az, Ellis, Falcon-Barroso, Flores, Gallazzi, Garc{\'\i}a-Lorenzo, Delgado, Gruel, Haines, Hao, Husemann, Iglesias-P{\'a}ramo, Jahnke, Johnson, Jungwiert, Kalinova, Kehrig, Kupko, Lopez-Sanchez, Lyubenova, Marino, Marmol-Queralto, Marquez, Masegosa, Meidt, Mendez-Abreu, Monreal-Ibero, Montijo, Mourao, Palacios-Navarro, Papaderos, Pasquali, Peletier, Perez, Perez, Quirrenbach, Rela{\~n}o, Rosales-Ortega, Roth, Ruiz-Lara, Sanchez-Blazquez, Sengupta, Singh, Stanishev, Trager, Vazdekis, Viironen, Wild, Zibetti, \& Ziegler}]{Sanchez2012CALIFAOverview}
Sanchez, S.~F., Kennicutt, R.~C., de~Paz, A.~G., {et~al.} 2012, \aap, 538, 31

\bibitem[{Sandage {et~al.}(1970)Sandage, Freeman, Stokes, Sandage, Freeman, \& Stokes}]{Sandage1970IntrinsicFlatteningESOSpirals}
Sandage, A., Freeman, K.~C., Stokes, N.~R., {et~al.} 1970, ApJ, 160, 831

\bibitem[{Sande {et~al.}(2017)Sande, Bland-Hawthorn, Brough, Croom, Cortese, Foster, Scott, Bryant, d'Eugenio, Tonini, Goodwin, Konstantopoulos, Lawrence, Medling, Owers, Richards, Schaefer, \& Yi}]{vandeSande2017SAMISlowRotators}
Sande, J. v.~d., Bland-Hawthorn, J., Brough, S., {et~al.} 2017, Monthly Notices of the Royal Astronomical Society, 472, 1272

\bibitem[{Sande {et~al.}(2019)Sande, Lagos, Welker, Bland-Hawthorn, Schulze, Remus, Bahe, Brough, Bryant, Cortese, Croom, Devriendt, Dubois, Goodwin, Konstantopoulos, Lawrence, Medling, Pichon, Richards, Sanchez, Scott, \& Sweet}]{vandeSande2019SAMISimulations}
Sande, J. v.~d., Lagos, C.~D., Welker, C., {et~al.} 2019, Monthly Notices of the Royal Astronomical Society, 484, 869

\bibitem[{{Schaye} {et~al.}(2015){Schaye}, {Crain}, {Bower}, {Furlong}, {Schaller}, {Theuns}, {Dalla Vecchia}, {Frenk}, {McCarthy}, {Helly}, {Jenkins}, {Rosas-Guevara}, {White}, {Baes}, {Booth}, {Camps}, {Navarro}, {Qu}, {Rahmati}, {Sawala}, {Thomas}, \& {Trayford}}]{Schaye2015EAGLE}
{Schaye}, J., {Crain}, R.~A., {Bower}, R.~G., {et~al.} 2015, \mnras, 446, 521

\bibitem[{{S{\'e}rsic}(1963)}]{Sersic1963Influencegalaxy}
{S{\'e}rsic}, J.~L. 1963, Boletin de la Asociacion Argentina de Astronomia La Plata Argentina, 6, 41

\bibitem[{Statler(1994)}]{Statler1994UncoveringIntrinsicShapes}
Statler, T.~S. 1994, ApJ, 425, 458

\bibitem[{Stein(1987)}]{Stein1987LargeSampling}
Stein, M. 1987, Technometrics, 29, 143

\bibitem[{van~de Sande {et~al.}(2017)van~de Sande, Bland-Hawthorn, Fogarty, Cortese, d'Eugenio, Croom, Scott, Allen, Brough, Bryant, Cecil, Colless, Couch, Davies, Elahi, Foster, Goldstein, Goodwin, Groves, Ho, Jeong, Jones, Konstantopoulos, Lawrence, Leslie, L{\'{o}}pez-S{\'{a}}nchez, McDermid, McElroy, Medling, Oh, Owers, Richards, Schaefer, Sharp, Sweet, Taranu, Tonini, Walcher, \& Yi}]{vandeSande2017SAMIHigherOrderKinematics}
van~de Sande, J., Bland-Hawthorn, J., Fogarty, L. M.~R., {et~al.} 2017, The Astrophysical Journal, 835, 104

\bibitem[{van~de Sande {et~al.}(2018)van~de Sande, Scott, Bland-Hawthorn, Brough, Bryant, Colless, Cortese, Croom, Deugenio, Foster, Goodwin, Konstantopoulos, Lawrence, McDermid, Medling, Owers, Richards, \& Sharp}]{vandeSande2018StellarAgesShapes}
van~de Sande, J., Scott, N., Bland-Hawthorn, J., {et~al.} 2018, Nature Astronomy, 483

\bibitem[{{van den Bosch} \& {van de Ven}(2009)}]{vandenBosch2009}
{van den Bosch}, R. C.~E., \& {van de Ven}, G. 2009, \mnras, 398, 1117

\bibitem[{{Villaescusa-Navarro} {et~al.}(2021){Villaescusa-Navarro}, {Angl{\'e}s-Alc{\'a}zar}, {Genel}, {Spergel}, {Somerville}, {Dave}, {Pillepich}, {Hernquist}, {Nelson}, {Torrey}, {Narayanan}, {Li}, {Philcox}, {La Torre}, {Maria Delgado}, {Ho}, {Hassan}, {Burkhart}, {Wadekar}, {Battaglia}, {Contardo}, \& {Bryan}}]{VillaescusaNavarro+:2010}
{Villaescusa-Navarro}, F., {Angl{\'e}s-Alc{\'a}zar}, D., {Genel}, S., {et~al.} 2021, \apj, 915, 71

\bibitem[{Vincent \& Ryden(2005)}]{Vincent2005ShapeLuminosityProfile}
Vincent, R.~A., \& Ryden, B.~S. 2005, ApJ, 623, 137

\bibitem[{Weijmans {et~al.}(2014)Weijmans, Zeeuw, Emsellem, Krajnovi{\'c}, Lablanche, Alatalo, Blitz, Bois, Bournaud, Bureau, Cappellari, Crocker, Davies, Davis, Duc, Khochfar, Kuntschner, McDermid, Morganti, Naab, Oosterloo, Sarzi, Scott, Serra, Kleijn, \& Young}]{Weijmans2014ShapesOfEarlyTypes}
Weijmans, A.~M., Zeeuw, P. T.~D., Emsellem, E., {et~al.} 2014, Monthly Notices of the Royal Astronomical Society, doi:10.1093/mnras/stu1603

\bibitem[{Wilkinson {et~al.}(2023)Wilkinson, Ludlow, Lagos, Fall, Schaye, \& Obreschkow}]{Wilkinson2023impactdiscs}
Wilkinson, M.~J., Ludlow, A.~D., Lagos, C. d.~P., {et~al.} 2023, Monthly Notices of the Royal Astronomical Society, 519, 5942

\end{thebibliography}

\appendix
\section{Galaxy Shape Predictions without Feature Selection} \label{apd:predwofs}

\begin{figure}
\centering
\includegraphics[width=0.95\linewidth,keepaspectratio]{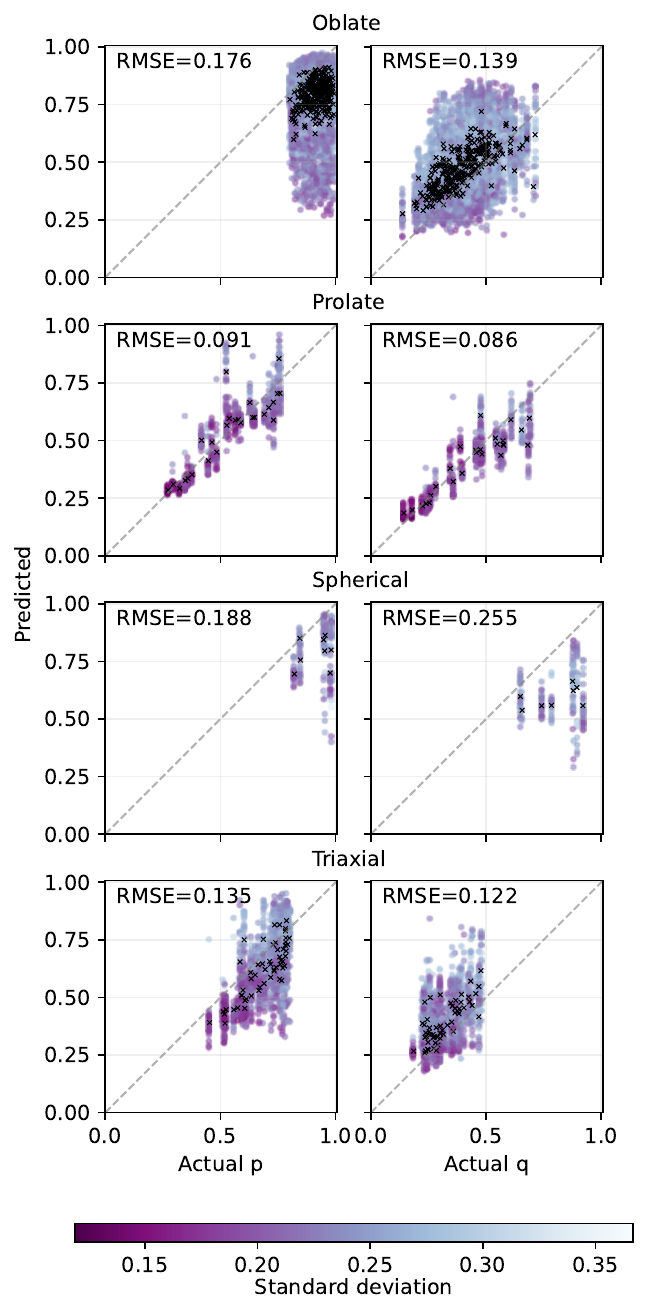}
\caption{Predicted against actual $p$ and $q$ for each galaxy shape using mixture density network (MDN) without performing feature selection for the test data set. The black crosses represent the average prediction, while circles represent projections of individual galaxies colour coded by the standard deviation from the MDN output. The darker the gradient, the more certain. The prediction error is evaluated by the root mean squared error (RMSE), where lower values represent better agreement. For reference, the identity is shown as a grey dashed line. Compared to Figure~\ref{fig:predvstrue_galclass}, the RMSE is marginally worse for all systems except the $p$ values for triaxial when no feature selection is performed.}
\label{fig:predvstrue_galclass_wopca}
\end{figure}

Figure~\ref{fig:predvstrue_galclass_wopca} shows the predicted mean of $p$ and $q$ from the MDN model when no feature selection is applied, i.e., using all parameters listed in Table~\ref{tab:features}.
In all cases except for $p$ values for triaxial systems, the MDN performs slightly better with lower RMSE when feature selection is included (see Figure~\ref{fig:predvstrue_galclass}).
Incorporating feature selection will be particularly useful especially in the presence of large number of features, which we intend to explore in future work.

\section{Trends in Certainty of Galaxy Shape Predictions} \label{apd:trendsigma}

\begin{figure*}
\centering
\includegraphics[width=0.49\columnwidth]{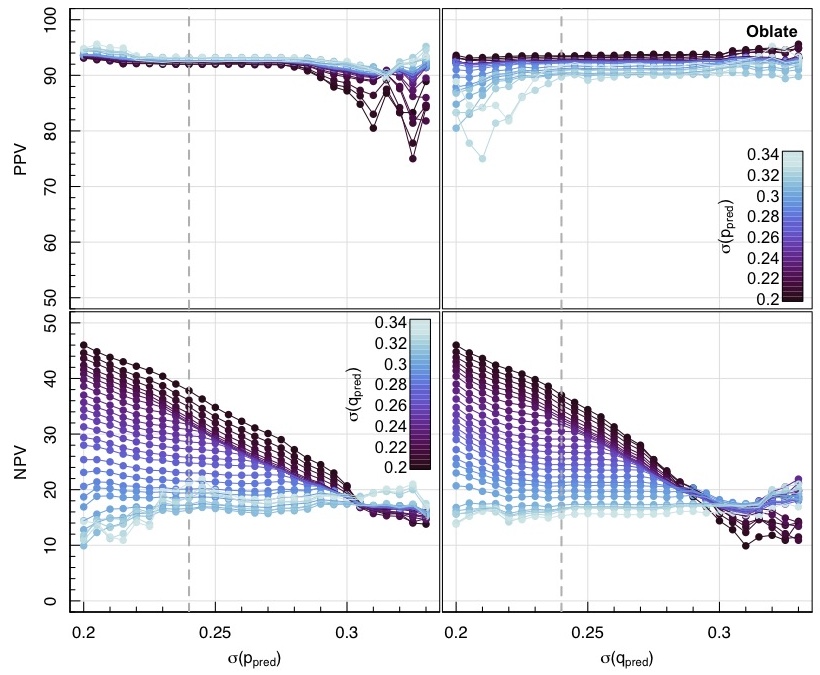}%
\includegraphics[width=0.49\columnwidth]{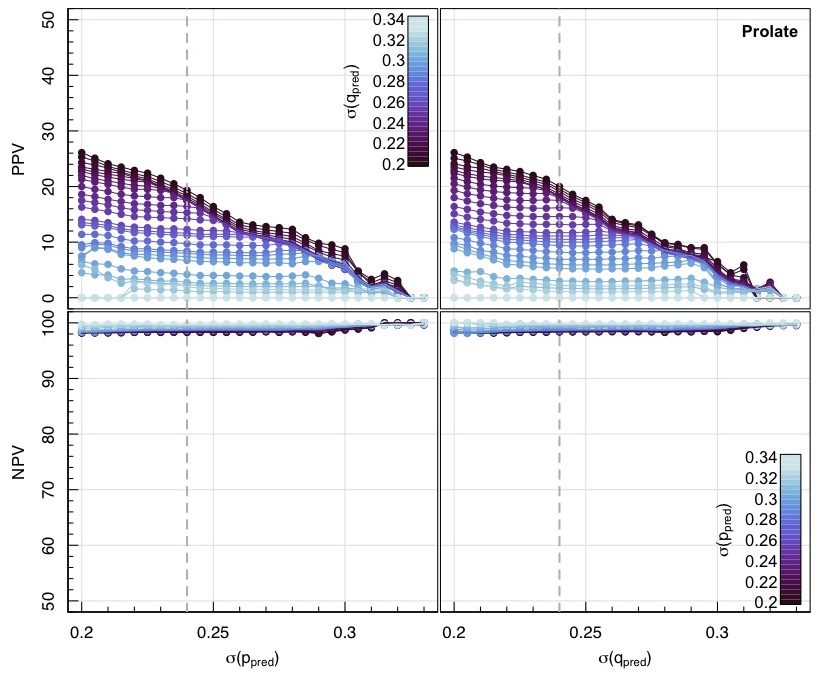}
\includegraphics[width=0.49\columnwidth]{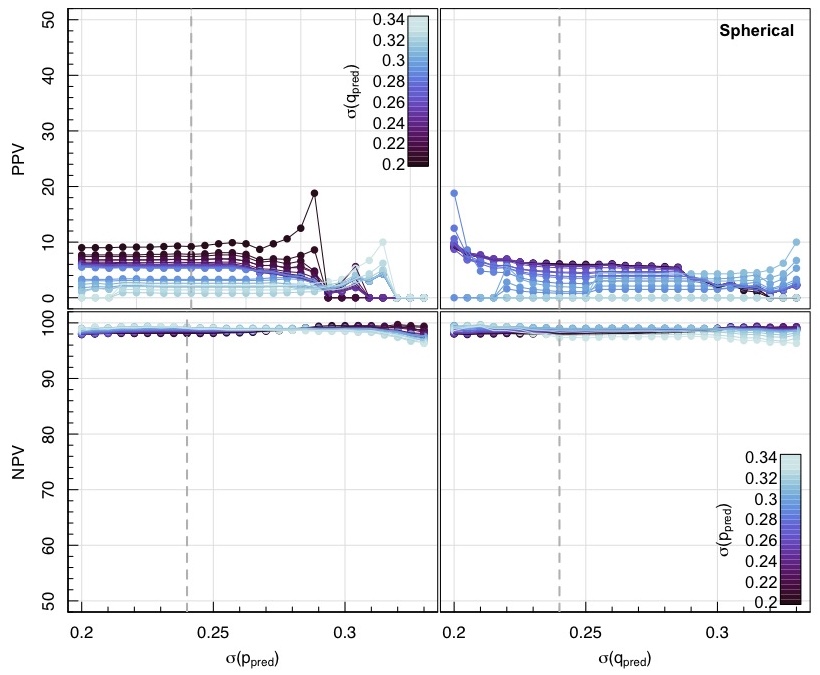}%
\includegraphics[width=0.49\columnwidth]{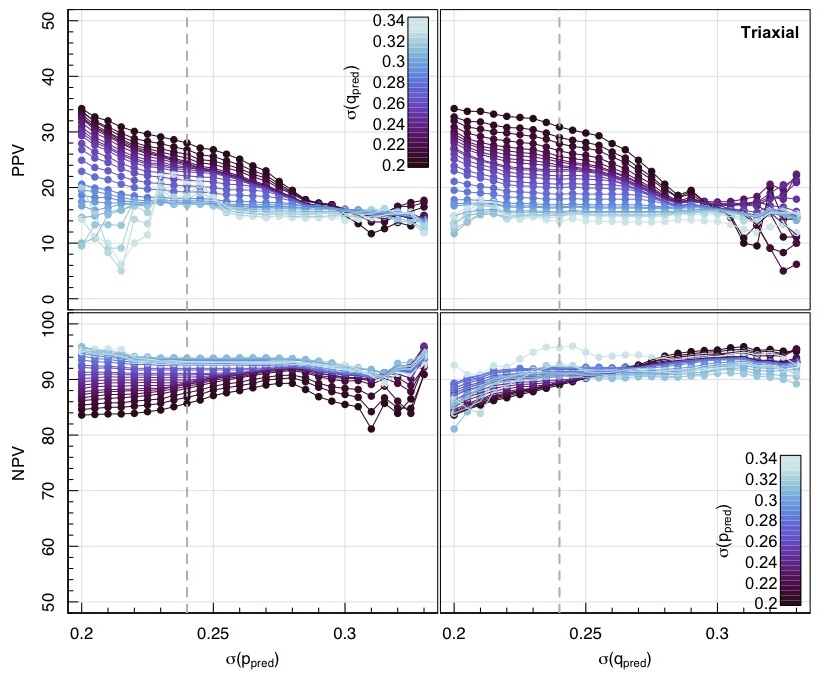}
\caption{Trends in positive predicted values (PPV) and negative predicted values (NPV) with varying standard deviations from MDN output ($\sigma$) for each galaxy shape. In each panel, the colour of the points shows the constant bin in the $p$ or $q$ value not on the axis, i.e., $\sigma(p_{\text{pred}})$ in the plot of $\sigma(q_{\text{pred}})$ and vice versa. The darker the gradient, the narrower the $\sigma$ and more certain. The vertical grey dashed line shows the mean of the standard deviation at $\sigma=0.24$.}
\label{fig:pred_gal_trends}
\end{figure*}

Figure \ref{fig:pred_gal_trends} shows the trends in positive predicted values (PPV) and negative predicted values (NPV) with different cut-off selection for the standard deviation of the MDN output.
The cut-off criterion for Figures \ref{fig:pdftrue_galclass}--\ref{fig:predp+q_galclass} and Table \ref{tab:results_stats} of the main paper using the mean of the standard deviation at $\sigma=0.24$ is shown in vertical grey dashed line.
Broadly, there are improvements in PPV and NPV as more certain objects are selected.

Notably for oblate systems, the NPV rate is refined by a factor of $\sim 4$ improvement.
The PPV for prolate and triaxial systems also improve by a factor of $\sim 2$ with the removal of less confident predictions.
The NPV results for triaxial systems behave opposite from expectation, yielding worse predictions as the standard deviation decreases.
This may be due to the expected distribution not being well captured by the MDN model, particularly for $p$ (see Figure \ref{fig:hist_galclass}).
The recovered $p$ values for triaxial objects cover a broader range of values compared to the true underlying shapes.
There is no improvement in the NPV for spherical systems due to the smaller sample size in that group.
In general, predictions with lower $\sigma$ are to be preferred to ensure the retrieved galaxy shapes are ``informative''.

\end{document}